\documentclass[twocolumn]{aastex7}
\usepackage{kotex}
\usepackage{multirow}
\usepackage{gensymb}
\usepackage{amsmath}
\usepackage{graphicx}
\usepackage{xcolor}
\usepackage[T1]{fontenc} 
\usepackage{enumitem}
\usepackage{comment}
\shorttitle{The XRISM/Resolve Spectrum of NGC~4151}
\shortauthors{J. M. Miller et al.}

\begin{document}

\title{The 0.9~Megasecond XRISM/Resolve Spectrum of the Seyfert-1 AGN NGC 4151}

\author[orcid=0000-0003-2869-7682]{Jon M. Miller}
\affiliation{Department of Astronomy, University of Michigan, 1085 South University Avenue, Ann Arbor MI 48109, USA}
\email[show]{jonmm@umich.edu}

\author[0000-0002-7129-4654]{Xin Xiang}
\affiliation{Department of Astronomy, University of Michigan, 1085 South University Avenue, Ann Arbor MI 48109, USA}
\email{xinxiang@umich.edu}

\author[0000-0002-4992-4664]{Missagh Mehdipour}
\affiliation{Department of Astronomy, University of Michigan, 1085 South University Avenue, Ann Arbor MI 48109, USA}
\email{missagh@umich.edu}

\author[0000-0001-9911-7038]{Liyi Gu}
\affiliation{SRON Space Research Organization Netherlands, Niels Bohrweg 4, 2333CA Leiden, The Netherlands} 
\email{L.Gu@sron.nl}

\author[0000-0001-9735-4873]{Ehud Behar}
\affiliation{Department of Physics, Technion, Technion City, Haifa 3200003, Israel}
\email{behar@physics.technion.ac.il}

\author[0000-0003-2663-1954]{Laura Brenneman}
\affiliation{Center for Astrophysics | Harvard-Smithsonian, MA 02138, USA}
\email{lbrenneman@cfa.harvard.edu}

\author[orcid=0000-0002-3687-6552]{Doyee Byun (변도의)}
\affiliation{Department of Astronomy, University of Michigan, 1085 South University Avenue, Ann Arbor MI 48109, USA}
\email{doyeeb@umich.edu}

\author[0000-0001-8470-749X]{Elisa Costantini}
\affiliation{SRON Space Research Organization Netherlands, Niels Bohrweg 4, 2333CA Leiden, The Netherlands} 
\affiliation{Anton Pannekoek Institute for Astronomy, University of Amsterdam, Science Park 904, NL-1098 XH Amsterdam, The Netherlands}
\email{E.Costantini@sron.nl}

\author[0009-0006-4968-7108]{Luigi Gallo}
\affiliation{Department of Astronomy and Physics, Saint Mary's University, Nova Scotia B3H 3C3, Canada}
\email{lgallo@ap.smu.ca}

\author[0000-0003-3828-2448]{Javier A. Garc\'ia}
\affiliation{NASA / Goddard Space Flight Center, Greenbelt, MD 20771, USA}
\affiliation{California Institute of Technology, Pasadena, CA 91125, USA}
\email{javier.a.garciamartinez@nasa.gov}

\author[0000-0002-1094-3147]{Matteo Guainazzi}
\affiliation{European Space Agency (ESA), European Space Research and Technology Centre (ESTEC), 2200 AG Noordwijk, The Netherlands} 
\email{Matteo.Guainazzi@esa.int}

\author[0000-0003-4511-8427]{Peter Kosec}
\affiliation{Center for Astrophysics | Harvard-Smithsonian, MA 02138, USA}
\email{peter.kosec@cfa.harvard.edu}

\author[0000-0002-2933-048X]{Takeo Minezaki}
\affiliation{Institute of Astronomy, School of Science, University of Tokyo, 2-21-1 Osawa, Mitaka, Tokyo 181-0015, Japan}
\email{minezaki@ioa.s.u-tokyo.ac.jp}

\author[0009-0009-0439-1866]{Daiki Miura}
\affiliation{Department of Physics, Graduate School of Science, The University of Tokyo, 7-3-1 Hongo, Bunkyo-ku, Tokyo 113-0033, Japan}
\affiliation{Institute of Space and Astronautical Science (ISAS), Japan Aerospace Exploration Agency (JAXA), 3-1-1 Yoshinodai, Chuo-ku,
Sagamihara, Kanagawa 252-5210, Japan}
\email{dikmur0611@g.ecc.u-tokyo.ac.jp}

\author[0000-0002-8108-9179]{Stephane Paltani}
\affiliation{Department of Astronomy, University of Geneva, Versoix CH-1290, Switzerland} 
\email{Stephane.Paltani@unige.ch }

\author[0000-0002-0572-9613]{Abderahmen Zoghbi}
\affiliation{Department of Astronomy, The University of Maryland, College Park, MD 20742, USA}
\affiliation{HEASARC, Code 6601, NASA/GSFC, Greenbelt, MD 20771, USA}
\affiliation{CRESST II, NASA Goddard Space Flight Center, Greenbelt, MD 20771, USA}
\email{azoghbi@umd.edu}

\begin{abstract}
NGC 4151 is the brightest Seyfert-1 active galaxy in the pass band of the Resolve calorimeter spectrometer aboard XRISM.  It has been observed on 14 occasions, resulting in a total exposure of 893~ks.  Herein, we report on an analysis of the time-averaged spectrum.  The narrow Fe~K$_{\alpha}$ emission line complex requires contributions from the torus and the optical broad line region (BLR).  Models assuming an emissivity index of $q=2$ for these components are statistically preferred over models assuming $q=3$ for a flat disk (where $J\propto r^{-q}$), consistent with a ``bowl'' geometry for the broader central engine.  A smooth shoulder on the red wing of these line components is likely best interpreted as Compton scattering in a medium with bound electrons, potentially signaling the presence of dust at the base of the BLR and in the torus.  The data also statistically prefer the addition of relativistic reflection from the innermost accretion disk, extending down to a radius of $r = 3.2^{+3.5}_{-2.0}~GM/c^{2}$ and with an inclination of $\theta = 29.7^{+0.5}_{-0.4}$~degrees.  The Fe K edge at 7.1~keV is best modeled with contributions from multiple charge states, consistent with obscuration due to cool, $kT \simeq 5$~eV collisional gas or photoionized gas rather than cold, neutral gas.  Dust is not evident in the Fe~K absorption edge.  A spectrum of outflows is clearly revealed, with slow ``warm absorber'' winds spanning Fe XX-XXVI, fast winds primarily seen via Fe~XXV and Fe~XXVI lines, and ultra-fast outflows (or, UFOs) seen as broad Fe~XXVI lines.  The warm absorbers are almost certainly ``failed'' winds that return to the central engine; the data constrain their radius, density, filling factor, and distribution.  For the most conservative volume filling factors, the UFOs may not deliver the kinetic feedback needed to halt star formation, on average.  However, they may generate galaxy-altering feedback for larger filling factors and/or during certain intervals.  We discuss these results in terms of the geometry of accretion flow in active galaxies, accretion--driven feedback into host galaxies, and future modeling efforts.
\end{abstract}


\section{Introduction}
Studies of active galactic nuclei take innumerable shapes, but many observations are now aimed at understanding how black holes grow, evolve, and influence their host environment.  This focus crystallized after cosmological simulations required AGN feedback to reproduce the evolution of host galaxies, particularly the prevention of subsequent star formation in host bulges (e.g. \citealt{dimatteo2005}, \citealt{hopkins2010}).  

Unfortunately, the orders of magnitude that span from the luminous innermost accretion disk to wind launching regions cannot be imaged in the UV and X-ray wavelengths that naturally characterize them.  X-ray spectroscopy and variability studies may provide the most direct view of the central engine, since X-ray light more readily penetrates gas and dust.  Moreover, most of the mass flux in accretion-driven winds lies at high ionization (e.g., \citealt{crenshaw2012}, \citealt{king2013}).  

An apparently neutral Fe~K$_{\alpha}$ emission line at 6.40~keV is the most prominent atomic feature in the X-ray spectrum of every active galactic nucleus (AGN).  This is due to the relatively high cosmic abundance of Fe, the high fluorescence yield of its low charge states, the modest product of these parameters for other elements that contribute lines close to 6.40~keV, and the prevalence of cold gas in the accretion flow (for a recent review, see \citealt{gallo2023}).  This combination of parameters should also make Fe~K$_{\alpha}$ lines prominent across a broad range in redshift; currently, this expectation is realized out to $z = 2.9$ \citep{ferugio2011}, and possibly higher redshifts \citep{iwasawa2012}.  

A clear mapping of this line to geometries within the accretion flow may allow black hole masses to be measured via reverberation techniques (for a review, see \citealt{cackett2021}), and eventually via single-epoch line widths in future surveys (e.g., via deep-field imaging with the NewAthena X-ray Integral Field Unit, see \citealt{barret2023}).  The first step is to develop the necessary mapping between line flux components and accretion flow geometries using local AGN.  

Observations with the High Energy Transmission Gratings (HETG) aboard Chandra made important progress in this regard.  In a survey of bright Seyfert galaxies, \cite{2010ApJS..187..581S} found that the Fe~K$_{\alpha}$ line in many sources has a width of $few \times 10^{3}~{\rm km}~{\rm s}^{-1}$, similar to the optical broad line region (BLR). This was an important indication that at least part of the Fe~K$_{\alpha}$ line might respond to an ionizing continuum -- as per, e.g., H$\beta$ lines -- and might eventually enable masses via reverberation techniques.  In their survey, \cite{2010ApJS..187..581S} did not find a compelling correlation between the widths of Fe~K$_{\alpha}$ and H$\beta$ lines, but the comparison did not rely on simultaneous nor lag-corrected data.

Among Seyfert-1 galaxies, NGC~4151 delivers the highest continuum flux and highest Fe~K$_{\alpha}$ line flux.  Reverberation in the H$\beta$ line gives a mass of $M_{BH} = 3.4\pm 0.4\times 10^{7}~M_{\odot}$ \citep{bentz2015}.  Using stacked Chandra/HETG spectra, \cite{Miller_2018} found evidence of asymmetry in the Fe~K$_{\alpha}$ line in NGC~4151, requiring emission from $r \simeq 1000~GM/c^{2}$ when fit with models that allow for special relativistic effects and weak gravitational redshifts.  This radius is suggestive of an inner extension to the optical BLR.  

Combining numerous observations made using XMM-Newton and Suzaku, \cite{Zoghbi_2019} found that the Fe~K$_{\alpha}$ line flux in NGC~4151 lags the X-ray continuum by $\tau = 3.3^{+1.8}_{-0.7}$ days, about half of the typical H$\beta$ lag (e.g., \citealt{bentz2015}).  This lag value formally agrees with the radii inferred via Chandra spectroscopy, and implies a black hole mass of $M_{BH} = 1.8^{+2.2}_{-1.1}\times 10^{7}~M_{\odot}$ for a geometric correction factor of $f = 4.13\pm 1.05$ \citep{grier2013}.  A value consistent with optically determined masses -- roughly twice higher -- would require a proportionately higher value of $f$ and might imply geometric differences between the line production regions (recall that $M_{BH} \propto fc\tau v^2 / G$, where $\tau$ is the measured lag between the continuum and responding line, and $v$ is the width of the responding line).  Reverberation in the Fe~K$_{\alpha}$ line has also been reported in NGC~3516 \citep{noda2023}, which is nearly as bright as NGC~4151.

\cite{2024ApJ...973L..25X} reported on the nature of the Fe~K$_{\alpha}$ emission line complex in the first two XRISM/Resolve spectra of NGC~4151 obtained during its Performance \& Verification phase.  The observed line profiles required (1) multiple velocity components, (2) models that explicitly include key atomic physics details (e.g., K$_{\alpha,1}$ and K$_{\alpha,2}$ lines in the appropriate 2:1 flux ratio, scattering shoulders, and more), and (3) physical convolution models to account for dynamical broadening where relativistic effects contribute subtly.  

Formal agreement between the radii inferred via IR reverberation mapping and interferometry (e.g., \citealt{lyu2021}, \citealt{grav20}) and the narrowest Fe~K$_{\alpha}$ component, and between optical and UV line widths and the intermediate Fe~K$_{\alpha}$ component (e.g., \citealt{bentz2015}), clearly tie parts of the line flux to the molecular torus and optical BLR.  Resolve spectra from a number of AGN appear to confirm this picture (see Figure 5 in \citealt{miller279}; also see \citealt{mehdipour2025} and \citealt{li2025}).  However, the broadest component in NGC~4151 could not be understood in terms of known geometries, and might nominally require a warp or other geometry at $r \simeq 100~GM/c^{2}$. 
 
The nature of the line component corresponding to $r \simeq 100~GM/c^{2}$ is just one of several questions that naturally flow from the initial analysis of the Fe~K$_{\alpha}$ lines in NGC 4151.  This component models broadened, asymmetric emission that smoothly extends from 6.4~keV to approximately 6.1~keV (see Figure 1 in \citealt{2024ApJ...973L..25X}).  It was not attributed to Compton scattering of photons from the narrow line core, owing to the absence of a sharp shoulder at 6.25~keV (expected from 180-degree scatters shifting line photons by 150~eV in a gas of free electrons; see, e.g, \citealt{gallo2023}).  More broadly, fits to the Fe~K$_{\alpha}$ line components preferred an emissivity of $q = 3$ (where $J \propto r^{-q}$) as per a flat accretion disk.  This is at odds with emission lines that connect to specific radii, requiring more solid angle per radius than a flat disk would contribute.  It is possible that a preference for $q = 3$ indicates only modest local departures from the broader disk profile.  Last, uncertainties surrounding the emissivity index complicate interpretations of the measured inclination parameters, which may give vital insights into the geometry of the accretion flow.  Specifically, the measured inclinations ($\theta=$10--34~deg.) are likely too low to reflect the angle at which the entire accretion flow is viewed, and may only reflect the inclination of the local geometry relative to the line of sight \citep{2024ApJ...973L..25X}.   

\cite{xiang2025} analyzed the variable and highly ionized wind absorption lines imprinted on all five XRISM/Resolve spectra of NGC~4151 obtained during the Performance \& Verification phase.  Complex combinations of ultra-fast outflows (UFOs; $v \simeq 0.03-0.3c$), very fast outflows (VFOs; $v \simeq 1000-9000~{\rm km}~{\rm s}^{-1}$), and warm absorbers ($v \leq 1000~{\rm km}~{\rm s}^{-1}$) were simultaneously detected at high levels of statistical confidence.  While low-ionization warm absorbers were readily detected in NGC~4151 and other AGN in Chandra/HETG spectra (\citealt{kraemer2005}, \citealt{couto2016}), VFOs had not been observed, and UFOs were observed as tentative, single-bin absorption features in CCD spectra (e.g., \citealt{tombesi2010}).

UFOs are more commonly associated with black holes that are likely accreting at or above the Eddington limit, notably PDS~456 \citep{pds456}.  The fact that UFOs are launched in NGC~4151 -- which accretes at just $\lambda_{Edd} \simeq 0.04$ -- signals that galaxy-altering feedback may persist over a greater fraction of an AGN lifetime (also see \citealt{zak2024}).  However, the flows seen in PDS~456 and NGC~4151 may be different.  Whereas the UFOs in PDS~456 have remarkably small velocity widths -- suggestive of clumps accelerated in a uniform way -- the fastest UFOs in NGC~4151 are fully resolved and fairly broad (e.g., bulk shifts of $v \simeq 0.15c \simeq 45\times 10^{3}~{\rm km}~{\rm s}^{-1}$, and widths of $\sigma = 5\pm 2\times 10^{3}~{\rm km}~{\rm s}^{-1}$).  The UFOs in NGC~4151 are likely launched and/or accelerated over a broad range of radii. 

The absorption measure distribution derived in the Resolve spectra of NGC~4151 is consistent with wind production and acceleration over a broad range of radii (${\rm AMD} = d N_{H}/d{\rm log}\xi$, where $\xi = L/nr^{2}$ is the ionization parameter with units of ${\rm erg}~{\rm cm}~{\rm s}^{-1}$; see \citealt{behar2009}).  This is consistent with magnetocentrifugal driving \citep{bp82}, which is expected to operate over decades in radius.  Moreover, estimates of the momentum rates in the radiation and winds suggest that radiation pressure cannot drive the wind.  

However, even these results leave open some questions, and also pose new ones.  The duty cycle and conditions that spur UFO production must be understood in order to assess whether kinetic feedback is likely to alter the host bulge in NGC~4151 and other Seyfert galaxies.  The gas densities within the differing wind types -- and therefore their volume filling factors and maximum kinetic luminosity -- are poorly constrained.  And, if warm absorbers are ``failed'' winds that do not escape to infinity, how far from the central engine do they reach?  

Herein, we present an analysis of the summed spectrum of NGC~4151, obtained from adding all 14 observations obtained during the Performance \& Verification phase and Cycle 1.  The total exposure of this spectrum is 893~ks.  Our analysis resolves some of the outstanding issues surrounding the nature of the Fe~K$_{\alpha}$ emission line complex and wind components, and how they trace the accretion flow geometry.  Separately, our analysis provides new insights into how the nature of variable ``neutral'' partial covering absorption within AGN, and the role of dust within the central engine.  

This approach carries at least two important caveats. First, a summed spectrum will necessarily emphasize narrow and less-variable features that are common to all individual observations, and may dilute variable features like UFOs.  Second, the summation of continua that might individually require different power-law indices -- not merely different normalizations -- could artificially create the appearance of relativistic reflection or other broad-band effects.  The time variability of the Fe~K$_{\alpha}$ emission line complex and the wind components will be treated in dedicated companion papers.

The details of the individual XRISM observations are given in Section 2.  Section 3 describes how the data were summed and analyzed.  The results of from our best-fit model and several alternatives are presented in Section 4.  In Section 5, we discuss these results, highlighting how they may impact our understanding of the accretion flow geometry and wind feedback in NGC~4151 and other AGN.  Finally, we restate our conclusions in Section 6. 

\section{Observation and Data Reduction}
\label{sec:observation}
We have utilized the Resolve \citep{ishisaki2022} data from all 14 XRISM observations of NGC 4151.  Five of these were obtained during the Performance and Verification phase, while nine were obtained during Cycle 1.  The observation identification numbers (ObsIDs), start times (in MJD), and net exposure of each observation are listed in Table 1.  Analysis of the Xtend data \citep{hayashida2018}, especially for long-term variablity, is deferred to a later paper.

In each observation, Resolve was run in its default mode, ``PX\_NORMAL''.  Following the XRISM Quick Start Guide version 2.3, and using HEASOFT version 6.34 and the associated calibration files, we filtered the data to avoid anomalies in low-Earth orbit, and to exclude specific pixels (pixel 36 is the calibration pixel; pixels 11 and 27 sometimes give anomalous readings).  The data were additionally filtered to only include high-resolution primary events (``Hp'' events), ensuring a resolution of 4.5~eV \citep{2024ApJ...973L..25X}.  We constructed light curves and spectra using standard procedures, and made response files using the tools \texttt{rslmkrmf} and \texttt{xaarfgen}.  We opted to construct ``Large'' redistribution matrix files rather than ``Extra Large'' versions for speed in spectral fitting.  The spectra and responses were added using the tool \texttt{ftaddpha}.  We verified its performance by adding subsets of the observations, and checking the results.  

Figure 1 shows the light curve of each observation over the 2.2--17.4~keV band, in a series of consecutive panels.  Note that the first five observations were not spaced evenly in time, whereas the last nine observations followed an every-other-day cadence.  While there is variability between and within the observations, they do not clearly group into the two states familiar in NGC~4151 (see, e.g., \citealt{Miller_2018}).  

\section{Analysis and Results}

\subsection{Set-up}

The data were analyzed using SPEX version 3.08.02 \citep{kaastra1996}. Prior to spectral fitting, the data were binned using the ``optimal'' binning algorithm of \cite{kaastra2016}.  The fits minimized a Cash statistic \citep{1979ApJ...228..939C}.  All of the errors reported in this work reflect parameter values at the boundary of their $1\sigma$ confidence intervals.

During the Performance and Verification Phase and Cycle 1, the gate valve was stuck in a closed position.  This nominally truncates the spectrum at 1.6~keV.  We adopt a lower fitting bound of 2.2~keV to avoid calibration uncertainties close to the boundary, and distortions that might arise through the use of the ``Large'' matrices.  The source remains well above the background to the limit of the Resolve pass band, 17.4~keV, so our fits were extended to this limit.  Backgrounds were ignored as they are small in comparison to a source as bright as NGC 4151.  In broad terms, our model construction follows the set-up adopted in \cite{2024ApJ...973L..25X} and \cite{xiang2025}, but details follow below.  

\subsection{Model set-up}

We utilize a continuum consisting of blackbody and power-law components, assumed to originate in a disk and corona.  As per \cite{xiang2025}, the blackbody has a temperature of $kT = 0.0168$~keV ($T = 1.9\times 10^{5}$~K), derived from their fits to UV spectra obtained with the Hubble Space Telescope Imaging Spectrograph (STIS).  Its luminosity is set to give an approximate 10:1 flux ratio relative to the power-law over the ionizing band (0.0136--13.6~keV).  The temperature and flux of the blackbody are fixed in our fits, as these parameters cannot be constrained via the Resolve spectrum.  Adopting exactly the same blackbody parameters used in \cite{xiang2025} enables consistent comparisons between the ionization parameters in that work and this paper.  The power-law component is attenuated at low energy and high energy using two ``etau'' functions in SPEX, to avoid errors in photoionization calculations.  (The ``etau'' functions define the optical depth as $\tau = \tau_0 E^{a}$, and we assume values of $\tau_0 = 1.3605\times 10^{-2}$ and $a=-1$, and $\tau_0 = 3.3333\times 10^{-3}$ and $a = 1$.)  The power-law index, $\Gamma$, and flux normalization are allowed to float freely within the fits.  Absorption in the Milky Way is predicted to be just $N_{H} = 2.1\times 10^{20}~{\rm cm}^{-2}$ \citep{bekhti2016}, orders of magnitude below internal columns.  For this reason, it is neglected. 

Both narrow and broad absorption lines are modeled with ``pion'' (\citealt{miller2015}, \citealt{mehdipour2016}), with zones layered by outflow velocity (i.e. outer zones see the continuum emission {\em after} it has been attenuated by other zones along the line of sight).  This effectively assumes that their observed velocity reflects the escape velocity where they were launched.  In each absorption zone, the equivalent neutral hydrogen column density (${\rm N}_{\rm H}$), the log of the ionization parameter ($\xi = L_{ion}/nr^{2}$, where $n$ is the number density and $r$ is the distance from the ionizing source), the rms gas velocity within the zone ($\sigma$), and the bulk velocity shift of the gas ($v$) were allowed to vary.  The absorption covering factor ($f_{cov}$) and emission covering factor ($\Omega \leq 4\pi$) were set to unity and zero, respectively.  Proto-solar abundances as per \cite{lodders2009} were assumed in all zones (this is the default within SPEX).  Only one zone requires substantial re-emission from the gas that is primarily seen in absorption; this is modeled with a ``pion'' zone with the absorption covering factor set to zero ($f_{cov} = 0$) but with a geometric covering factor allowed to vary ($\Omega/4\pi \leq 1$).  Note that \cite{xiang2025} assume an absorption covering factor of $f_{cov} = 0.5$, introducing a small constant offset between the two analyses.

Relative to \cite{2024ApJ...973L..25X} and \cite{xiang2025}, we model the complex of prominent Fe~K$_{\alpha}$ and K$_{\beta}$ emission lines in an updated manner:

The prior work used ``mytorus'' line kernels \citep{murphy2009} modified by the ``speith'' blurring function \citep{speith1995}.  While suited to the Resolve data in all key respects, the ``mytorus'' kernel assumes that photons from the line core are scattered only by free electrons.  Since full 180-degree scatters cause $\Delta E = -150$~eV line shifts, a shoulder is predicted at 6.25~keV when the gas column density is sufficiently high.  The sharpness of this feature is reduced if Doppler broadening is important.  

The absence of a sharp shoulder in the Resolve spectra of NGC~4151 led the smooth, line-like flux between 6.1--6.4~keV to be modeled as a broad emission line component from $r \simeq 100~GM/c^{2}$.  Sharp shoulders now appear to also be absent in other Resolve spectra of Seyfert-1 AGN, and smooth red wings extending down as far as 6.1~keV may be seen (e.g., Mrk 279: \citealt{miller279}; NGC 3783: \citealt{mehdipour2025}, \citealt{li2025}; NGC 7213: \citealt{kammoun2025}; NGC 3516: \citealt{juranova2025}).  This scattering is most pronounced in the XRISM/Resolve spectrum of the Circinus Galaxy (Ueda et al.\ 2026, submitted).  The common character of these lines -- across sources with different Eddington fractions and viewing angles -- suggests a potential origin in atomic processes.

In a medium where some of the electrons are bound to gas and dust, Compton scattering will give rise to a smooth shoulder owing to the range of binding energies.  Especially if iron is bound to dust at the base of the optical BLR (e.g., \citealt{czerny2015}), these conditions may be natural to both the BLR and torus.  We therefore model the narrowest (torus) and intermediate (BLR) components of the summed spectrum of NGC~4151 with separate pairings of a ``speith'' convolution function acting on an ``XCLUMPY'' \citep{tanimoto2019} line kernel.  The ``XCLUMPY'' model is very similar to ``mytorus'' in terms of the atomic physics included to describe the line core, and also similar in that it self-consistently models associated Fe~K$_{\beta}$ lines with relative strengths set by atomic data, but differs in that it assumes scattering in a medium with bound electrons.  Using two instances of ``XCLUMPY'' is not fully self-consistent as this would nominally amount to partially overlapping toroidal geometries; how distinct the BLR and torus are in extent and composition is a matter that observations can test.

For each instance of ``speith,'' the inner disk radius and inclination are allowed to vary.  In separate fitting experiments with a full model, we examine fits with the emissivity index fixed at $q=2$ and $q = 3$.  Within each ``XCLUMPY'' component, the gas column density (${\rm N}_{\rm H}$) and normalization were allowed to vary.  The power-law index of the incident radiation was linked to the same parameter in the continuum power-law, and the angle at which the torus is observed was linked to the inclination within ``speith.''  The opening angle of the torus was fixed at 45~degrees.

In contrast to the individual spectra of NGC 4151, a very broad, potentially relativistic Fe~K line becomes apparent in the 0.9~Ms spectrum.  We model this with ``speith'' acting on a high-resolution version of the well-known ``xillver'' disk reflection model \citep{garcia2013}.  The ``xillver'' power-law index was linked to the same parameter in the continuum power-law, the iron abundance relative to solar is fixed at unity (${\rm A}_{\rm Fe} = 1$), the inclination is linked to the same parameter in the ``speith'' function, and the gas density was frozen at its default value of ${\rm log}n = 15$.  The reflection ionization and normalization are allowed to vary freely.  The ``speith'' inner radius and inclination parameters are allowed to float freely, while the emissivity index is fixed at $q = 3$ and the spin was fixed to $a = 0.9$ (where $a = cJ/GM^{2}$, following \citealt{keck2015}). 

Separately, \cite{2024ApJ...973L..25X} and \cite{xiang2025} model apparently neutral partial-covering absorption within the central engine of NGC~4151 in terms of cold, neutral gas.  This predicts a sharp edge at 7.112~keV that is at odds with individual Resolve spectra of NGC~4151 (see Figure 2 in \citealt{xiang2025}).  In the summed 0.9~Ms spectrum, the difference between the data and a model with a sharp, neutral Fe~K edge cannot be ignored.  The edge is fit much better when the partial covering absorber is allowed to be cool, so that edges from several charge states contribute. Partial covering absorption within the system is therefore modeled using a ``hot'' component, with the equivalent neutral hydrogen column density (${\rm N}_{\rm H}$), covering factor ($f_{cov}$), and gas temperature allowed to vary freely.  

The best-fit model is detailed in Table 2.  It can be written as follows:\\

\noindent $ [bb + pow*etau*etau + speith*xillver + 2\times(speith*xclumpy)]*pion_{UFO-1}*pion_{UFO-2}*pion_{VFO-1}*pion_{WA-1}*pion_{WA-2}*hot*reds$,\\

\noindent where the $bb$ and $pow$ components are the blackbody and power-law continua, and the $etau$ components bend the power-law to zero at low and high energy values.  Note that only these direct continuum components feed into pion absorbers to define the ionizing luminosity, but the Fe~K$_{\alpha}$ emission line components (via ``xclumpy'' and ``xillver'') are also covered by the pion absorbing layers.  Again, the pion layers are ordered by outflow speed, assuming that faster winds have overcome a greater potential.  The misnamed ``hot'' component models cool, partial covering gas.  The ``reds'' component simply redshifts the full model to the frame of NGC~4151.

Finally, we note that we added narrow Gaussian functions to the phenomenologically model weak emission lines from Si, S, Ar, and Ca between 2--4~keV.  These are included merely to prevent the continuum fit from being biased; the lines likely originate in the extended X-ray narrow line region rather than the central accretion flow (see, e.g., \citealt{kraemer2005}, \citealt{wang2011}).  In later work, these lines will be utilized to characterize the abundances in the nuclear region of NGC~4151; they are not treated further in this work.

\section{Results}

The results of our fits are detailed in Table 2.  The model is necessarily incomplete, achieving a Cash statistic of $C = 5200$ for $\nu = 4848$ degrees of freedom.  It is a modest evolution from the models applied to individual observations that enables progress. The best-fit model is shown in Figure 2,  over the full pass band and again between 6.0--8.5~keV.  Figure 3 partially reproduces Figure 2, but shows how different emission components contribute to the broad-band flux, and the flux in the Fe~K band.  Figure 4 shows the best fits that result after (1) different emission components are removed from the total model, and (2) the total model is fit again so that other components can compensate.  The effects of removing specific absorption components and re-fitting are shown in Figure 5.  

In the sections below, we evaluate the significance of many components (or parameter choices) via changes in the Akaike Information Criterion (or, AIC; see \citealt{Akaike_1974}).  We use the adapted form utilized in \cite{Emmanoulopoulos_2016} and \cite{xiang2025}.  Models with or without a given component provide equally good fits for $\Delta{\rm AIC} > -2$ (the model with a given component should have a lower AIC).  For $\Delta{\rm AIC} < -10$, the inclusion of a given component is strongly supported.

\subsection{The broad line region and torus}

Confirming \cite{2024ApJ...973L..25X}, the narrow Fe~K$_{\alpha}$ (and, K$_{\beta}$) line complex unambiguously requires contributions from the BLR and torus (see Figures 2, 3, 4).  If the BLR component of the line is excluded, the best possible fit with the remaining components gives a Cash statistic of $C = 6112$, or $\Delta C = 912$.  If the torus component is instead excluded, the best possible fit with the remaining components gives $C = 5674$, or $\Delta C = 474$.  Both components are required at extremely high confidence levels.  The data strongly prefer an emissivity index of $q = 2$.  If a value of $q = 3$ is assumed for the BLR component, the fit statistic increases by $\Delta C = 102$.  Applying the Akaike Information Criterion, this represents a change of $\Delta{\rm AIC} = -100$, signaling that $q=2$ is very strongly perferred.  When $q = 3$ is assumed only for the torus component, the fit statistic increases by $\Delta C = 26$ for $\Delta \nu = 1$ (here, $\Delta{\rm AIC} = -26$, which is again highly significant).  This likely signals that the data are sensitive to the vertical extent of the BLR and torus relative to a flat disk described by an emissivity index of $q = 3$.  The measured preference for a $q = 2$ emissivity differs from the initial results reported in \cite{2024ApJ...973L..25X}; it is likely due to the different line kernel used in this analysis.

We find that the inner radius of the BLR component is $r_{BLR} = 1.9^{+1.3}_{-0.5}\times 10^{3}~GM/c^{2}$, and the inner radius of the torus component is $r_{tor} = 2.8^{+2.0}_{-1.0}\times 10^{3}~GM/c^{2}$.  Both radii are formally consistent with those initially measured in the first two Resolve spectra of NGC~4151 ($r_{BLR} = 1.9-3.3\times 10^{3}~GM/c^{2}$, $r_{tor} = 2.5-67\times 10^{3}~GM/c^{2}$).  The relatively large fractional errors on these radius values likely reflects the diminished sensitivity of the $q = 2$ emissivity profile ($q = 2$ implies that successive annuli with constant $\Delta r$ contribute as much to the line flux as interior radii; this is not the case if $q=3)$.  The implied radial separation between the BLR and torus is slightly smaller than the 0.5~dex separation suggested by IR reverberation mapping of the torus in quasars (e.g., \citealt{minezaki2019}), but 0.5~dex is not excluded.  

The inclination of the BLR component is measured to be $\theta = 48^{+38}_{-9}$~degrees.  In contrast, the inclination of the torus component is measured to be $\theta = 9\pm 1$~degrees.  This may indicate that the observed line flux is weighted by portion of the face of the torus that is normal to our line of sight.  This could be true if the torus subtends a larger solid angle than the BLR, as posited in AGN unification models \citep{antonucci1993}.  This is at least qualitatively consistent with a ``concave'' or ``bowl'' geometry for the accretion flow.

Our fits suggest that the observed gas is nearly optically thick, with a BLR column of $N_{H} = 1.4^{+7.6}_{-0.3}\times 10^{24}~{\rm cm}^{-2}$, and a torus column of $N_{H} = 0.4^{+0.5}_{-0.2}\times 10^{24}~{\rm cm}^{-2}$.  Especially given that the data appear to indicate that some electrons may be bound, these line components could arise in dusty gas.  The cross-sections of dust are strongly wavelength-dependent and a nominally optically thin column in X-rays may correspond to an optically thick column in UV (see, e.g., \citealt{wd2001}).  This is consistent with models of the BLR that invoke radiation pressure on dust to lift gas above the disk (e.g., \citealt{czerny2015}).

Our use of line kernels that assume bound electron scattering eliminates the need for emission from 
$r \simeq 100~GM/c^{2}$ in the spectrum of NGC 4151 initially suggested in \cite{2024ApJ...973L..25X} (see Figure 3).  Instead of dynamically broadened emission from a novel geometry, the smooth shoulder between 6.1--6.4~keV is modeled in terms of atomic scattering effects in the BLR and torus.  In a prior analysis of Chandra/HETG spectra of NGC~4151, a narrow annulus of emission at $r \simeq 100~GM/c^{2}$ was implied in difference spectra \citep{Miller_2018}; flux-dependent scattering -- perhaps mediated by a flux-dependent column density -- could potentially account for those results.

\subsection{The innermost disk}

The data strongly require broad curvature and complexity beyond what the continuum, narrow Fe~K$_{\alpha}$ emission line components, and absorption can produce (see Figure 3).  This is fit as blurred reflection from the innermost accretion disk, via $speith\times xillver$.  If this component is not included in the model, the best fit that can be produced by the other components gives a Cash statistic of $C = 7469$, or $\Delta C = 2263$ (see Figure 4).  We measure an inner disk radius of $r_{in} = 3.2^{+3.5}_{-2.0}~GM/c^{2}$, which nominally allows for both maximal and zero spin.   When the emissivity index, a break index, and the spin are allowed to vary, the fit is not improved and the spin cannot be constrained.  This may be partly due to the fact that the innermost radius within the ``speith'' model is not linked to the innermost stable circular orbit (ISCO) for a given spin value.  However, the fit also did not improve when corresponding pairings of inner radius and spin were fixed.

The measured inclination is very tightly constrained at $\theta = 29.7^{+0.5}_{-0.4}$~degrees.  If GRMHD effects anchor the disk in the plane defined by the black hole spin vector, then this value may indicate the inclination that defines the system.  This value is lower than the apparent inclination of the NLR bi-cone, $\theta = 45\pm 5$~degrees inferred via Hubble observations \citep{das2005}, but it is compatible with values measured by \cite{nandra2007} in an XMM-Newton survey of Seyferts ($\theta = 0-35$~deg., depending on the observation; also see \citealt{keck2015}).  

\subsection{Ultra-fast outflows}

Our best-fit model includes two UFOs.  The faster outflow (UFO-1 in Table 2) has an outflow velocity of $-v = 5.03\pm 0.12\times 10^{4}~{\rm km}~{\rm s}^{-1}$, or approximately $-v \simeq 0.17c$.  Unlike the UFOs in PDS~456 \citep{pds456}, this component has a substantial velocity width of $\sigma = 4.4^{+1.3}_{-0.9}\times 10^{3}~{\rm km}~{\rm s}^{-1}$, giving $\sigma/v \simeq 0.09$.  As seen in Figure 5, the main absorption component is the Fe~XXVI line shifted to approximately 8.2 keV, spanning between 8.0--8.4~keV.  The presence of this trough and the absence of a feature corresponding to Fe~XXV drives the model to a high ionization parameter, ${\rm log}\xi = 4.00^{+0.15}_{-0.09}$.  Of course, this range includes the energy of the Fe XXVI Ly-$\beta$ line (8.246~keV) at low velocity shifts; this is likely a coincidence, fostered by a broad UFO velocity with.

This UFO component was clearly detected in the spectra of NGC~4151 obtained during the Performance \& Verification phase.  In four of the five PV observations, \cite{xiang2025} detect a UFO with $-v = 4.5-4.8\times 10^{4}~{\rm km}~{\rm s}^{-1}$.  In those data, the column densities vary from non-detection to values as high as $N_{H} = 1.5\times 10^{23}~{\rm cm}^{-2}$.  The value that we have measured in the summed 0.9~Ms spectrum, $N_{H} = 2.2^{+1.1}_{-0.5}\times 10^{22}~{\rm cm}^{-2}$, may then represent an average.  Potentially, the properties that emerge from this UFO in the summed spectrum can be used to characterize a meaningful time-averaged mass outflow rate and kinetic power.

Starting from an assumption of spherical shells that follow an $n\propto r^{-2}$ density profile, a simple expression for the mass outflow rate in a wind component is $\dot{M} = \Omega \mu m_p n r^{2} v$ (see, e.g., \citealt{blustin2005}.  Here, $\Omega$ is the covering factor ($0 \leq \Omega \leq 4\pi$, and we assume a value of $\Omega = 2\pi$ because roughly 50\% of Seyfert-1 AGN have shown evidence of UFOs), $\mu$ is the mean atomic weight and we assume $\mu = 1.23$ as per cosmic abundances, $m_p$ is the proton mass, $n$ is the number density, and $v$ is the outflow velocity.  This can be rearranged using the ionization parameter formalism, giving $\dot{M} = \Omega \mu m_p (L_{ion}/\xi) {v} f_{v}$, where $f_{v}$ is an unknown volume filling factor.  The kinetic power in a given wind zone is then given by $L_{kin} = \frac{1}{2} \dot{M} v^{2}$.  The fact that $L_{kin} \propto v^{3}$ means that UFOs will dominate the total kinetic power budget in winds.

For UFO-1, $\dot{M} = 6.0^{+1.5}_{-0.9}\times 10^{26}~{\rm g}~{\rm s}^{-1}$, or $\dot{M} = 9.5^{+3.0}_{-2.4}~{M}_{\odot}~{\rm yr}^{-1}$, for volume filling factors of unity.  The kinetic power in this component is then $L_{kin} = 7.6^{+2.3}_{-2.5}\times 10^{45}~{\rm erg}~{\rm s}^{-1}$.  Taking $L_{bol} \simeq 2\times L_{ion}$, the mass accretion rate can be estimated by $\dot{M}_{acc} = L_{bol}/\eta c^{2} \simeq 1.0\times 10^{24}~{\rm g}~{\rm s}^{-1}$ (where $\eta$ is the accretion efficiency factor, and we assume $\eta = 0.1$).

These numbers nominally suggest that the mass outflow rate and kinetic luminosity of UFO-1 greatly exceed the mass accretion rate and total radiative luminosity, respectively.  However, \cite{xiang2025} note that the volume filling factor may be as small as $f_{v} = 5\times 10^{-4}$.  With this correction, $\dot{M} = 3.0^{+0.8}_{-0.9}\times 10^{23}~{\rm g}~{\rm s}^{-1} = 4.8^{+1.5}_{-1.2}\times 10^{-3}~{M}_{\odot}~{\rm yr}^{-1}$, and $L_{kin} = 3.8^{+1.2}_{-1.3}\times 10^{42}~{\rm erg}~{\rm s}^{-1}$.  This lower-limit mass outflow rate is comfortably lower than the inferred mass accretion rate.   \cite{dimatteo2005} and \cite{hopkins2010} find that wind can halt star formation in host galaxy bulges when the kinetic luminosity exceeds 0.5-5\% of the Eddington luminosity.  For a black hole mass of $M_{BH} = 3.4\times 10^{7}~M_{\odot}$ \citep{bentz2015}, $L_{Edd} = 4.4\times 10^{45}~{\rm erg}~{\rm s}^{-1}$.  For the smallest possible filling factor, $L_{kin}/L_{Edd} \simeq 8\times 10^{-4}$.

The second UFO in the time-averaged spectrum (UFO-2 in Table 2) has a lower velocity, $-v = 1.6\pm 0.1\times 10^{4}~{\rm km}~{\rm s}^{-1}$ or $-v \simeq 0.05c$.  The Fe~XXV resonance line within this component absorbs part of the neutral Fe~K$_{\beta}$ line (see Figure 5), and contributes a far weaker Fe XXVI absorption line at higher energy owing to its lower ionization (${\rm log}\xi = 3.5\pm 0.1$).  The character of this UFO is different than UFO-1, in that it has a low internal velocity of $\sigma = 160^{+60}_{-40}~{\rm km}~{\rm s}^{-1}$, giving $\sigma/v \simeq 0.01$.  It is possible that the two UFO components are not co-spatial within a broader outflow, and even possible that they are driven by different mechanisms.

There are plausible counterparts to UFO-2 among the wind components reported by \cite{xiang2025}.  In three of the five PV observations, a UFO is detected with $-v = 1.4-2.9\times 10^{4}~{\rm km}~{\rm s}^{-1}$ and ionization parameters between ${\rm log}\xi = 3.1-3.4$.  The column densities range from values that fall below detection limits, to $N_{H} = 0.7-70\times 10^{22}~{\rm cm}^{-2}$.  Here again, especially given the non-detections in two PV observations, the low column density measured in the summed 0.9~Ms spectrum -- $N_{H} = 1.1\pm 0.3\times 10^{21}~{\rm cm}^{-2}$ -- might represent a meaningful long-term average value.

Removing UFO-1 from the model and allowing all other components to compensate gives a fit with a Cash statistic of $C = 5262$, or $\Delta C = 56$, giving $\Delta{\rm AIC}=-54$.  Removing UFO-2 from the model and allowing all other components to compensate gives a fit with a Cash statistic of $C = 5258$ or $\Delta C = 52$, giving $\Delta{\rm AIC}=-50$.  The consequences for the fit are shown in Figure 5.  As shown in \cite{xiang2025}, the significance of these features is higher in some individual observations owing to their strong time variability.  

\subsection{Very fast outflows}

The spectrum strongly requires one very fast outflow (VFO-1 in Table 2).  This component has an outflow velocity of $-v = 3.6^{+0.2}_{-0.4}\times 10^{3}~{\rm km}~{\rm s}^{-1}$, and a velocity width of $\sigma = 1.1^{+0.5}_{-0.2}\times 10^{3}~{\rm km}~{\rm s}^{-1}$ (giving $\sigma/v \simeq 0.3$, much higher than the UFOs).  Its ionization parameter, ${\rm log}\xi = 3.21\pm 0.01$, is lower than the UFOs but lies in between the ionization parameters of the warm absorbers (see below).  This departure from a 1:1 correspondence between velocity and ionization is expected if multiple driving mechanisms contribute, and/or if clumping and circulation are important along the line of sight.  The column density of this wind component is modest, just $N_{H} = 6.3\pm 0.8\times 10^{21}~{\rm cm}^{-2}$, but it is clearly detected as a blue extension of the strong Fe~XXV resonance absorption line at 6.70~keV (see Figure 5).  Without a VFO component, the best-fit model for the data gives a Cash statistic of $C = 5355$, or $\Delta C = 155$, giving $\Delta{\rm AIC} = -147$.

\subsection{Warm absorbers}

The strong, narrow absorption lines between 6.5 and 7.0~keV can be modeled via pion zones with distinct ionization parameters but similar velocities (see ``WA-1'' and ``WA-2'' in Table 2).  The strong line at 6.70~keV and the doublet at 6.95 and 6.97~keV are  Fe XXV and Fe XXVI, respectively.  These lines require a high ionization (${\rm log}\xi = 3.75\pm 0.03$), a modest velocity broadening ($\sigma = 220\pm 10~{\rm km}~{\rm s}^{-1}$), and modest blue-shift ($-v = 290^{+10}_{-20}~{\rm km}~{\rm s}^{-1}$).  Here, $\sigma/v \simeq 0.8$, much higher than both the VFO and UFO components.  The series of lines seen between 6.50--6.70~keV that are consistent with Fe~XX--XXIV require a different ionization parameter, ${\rm log}\xi = 3.259^{+0.007}_{-0.006}$.  This presence of two distinct ionization parameters but roughly similar velocities is suggestive of a multi-phase gas.  

The column densities of these warm absorber components are modest, just $N_H = 2.0^{+0.1}_{-0.2}\times 10^{22}~{\rm cm}^{-2}$ and $N_H = 7.2^{+0.6}_{-0.4}\times 10^{21}~{\rm cm}^{-2}$, respectively.  However, the combination of their low velocity widths the and extraordinary sensitivity afforded by 0.9~Ms of exposure serves to make them prominent within the overall spectrum.  Note that $\Delta E = 4.5$~eV corresponds to $\sigma = 202~{\rm km}~{\rm s}^{-1}$ at 6.70~keV, so even the warm absorbers are marginally resolved.

The highly ionized warm absorber, WA-1, is observed with a P Cygni profile (see Table 2, and Figures 2, 3, and 4).  The emission component is modeled using a connected pion emitter, with its column density, ionization parameter, and internal velocity broadening linked to the values of the absorption component.  The bulk shift of the emission was allowed to vary freely, and the emission covering factor was allowed to vary in the range $0\leq \Omega/4\pi \leq 1$.  Dynamical broadening is captured by convolving this pion emitter with a Gaussian, via the ``vgau'' model (adding a single width parameter).  In total, then, the additional component is described by three additional parameters. 

The flux in this component models emission to the red of the Fe XXVI and Fe XXV absorption lines.  It also serves to partly fill-in the 6.95~keV component of the Fe~XXVI absorption doublet, explaining why the 6.95~keV feature is weaker than expected relative to the 6.97~keV line (see Figure 3).  Removing the emission component and allowing all other components to vary results in a Cash statistic of $C = 5396$, or $\Delta C = 196$ for two degrees of freedom ($\Delta{\rm AIC} = -192$).  

This emission component is red-shifted by $v = 870\pm 30~{\rm km}~{\rm s}^{-1}$, potentially indicating the full bulk velocity of the zone better than the projected velocity measured in absorption.  The width of the emission component is $\sigma = 970^{+160}_{-90}~{\rm km}~{\rm s}^{-1}$.  If this broadening is a Keplerian orbital speed, it implies a radius of $r = 9.6^{+2.0}_{-2.5}\times 10^{4}~GM/c^{2}$.  This warm absorber component is then observed at a greater distance from the central engine than the inner faces of the BLR and torus.  If the two warm absorbers are truly cospatial, then the less ionized component is located at similar radii.

Rewriting the ionization parameter formalism and assuming a filling factor of unity gives an upper limit on the absorption radius, $r \leq L_{ion}/N_{H}\xi$.  The best-fit model in Table 2 gives an ionizing luminosity of $L_{ion} = 9.2\times 10^{43}~{\rm erg}~{\rm s}^{-1}$.  This implies an absorption radius of $r \lesssim 8.2\times 10^{17}~{\rm cm}$ or $r \lesssim 1.6\times 10^{5}~GM/c^{2}$ for the highly ionized warm absorber, for a black hole mass of $M_{BH} = 3.4\times 10^{7}~M_{\odot}$ \citep{bentz2015}.  This is comparable to the radius derived by assuming that the broadening of the emission component is solely due to Keplerian motion, and suggests a high filling factor.

With the dynamical radius measurement, the ionization parameter can be rewritten to derive the gas density: $n = L_{ion}/r^{2}\xi \simeq 6.8^{+5.6}_{-2.2}\times 10^{4}~{\rm cm}^{-3}$.  This value is orders of magnitude below the density inferred in the optical BLR, where the absence of forbidden lines implies $n \geq 10^{8}~{\rm cm}^{-3}$.  The gas filling factor (clumping factor) can be estimated by calculating the length required to produce the observed column density and comparing it to the dynamical radius: ${\rm N}_{\rm H} = n \Delta r$ gives $\Delta r = 3.0\pm 1.3\times 10^{17}~{\rm cm}$, and $\Delta r/r \simeq 0.6$.  At least in the case of NGC~4151, the data indicate that highly ionized warm absorbers are not strongly clumped, with a filling factor of close to unity.  Moreover, our fits also suggest that the gas covers most of the solid angle, $\Omega/4\pi = 1.0_{-0.1}$.   These properties may partially explain why warm absorbers are so common, and even seen in some Seyfert-2 AGN (e.g., NGC 4388; see \citealt{miller2019}, \citealt{gediman2024}).

It is now also possible to assess whether the warm absorbers escape to infinity, or remain bound.  The more highly ionized warm absorber, at least, may be launched with a velocity of at least $v = 870\pm 30~{\rm km}~{\rm s}^{-1}$.  This is the escape speed from $r \simeq 2.3\times 10^{5}~GM/c^{2}$.  However, we infer that the wind is launched from $r = 9.6^{+2.0}_{-2.5}\times 10^{4}~GM/c^{2}$, roughly twice as small. The warm absorber components likely represent failed winds that do not escape from the central engine.  

\cite{xiang2025} found that the winds in the five XRISM PV observations of NGC~4151 may be largely magnetocentrifugal, based on the absorption measure distribution of the gas.  Such winds are expected to be launched at an angle of $\theta = 60$~deg. with respect to the plane of the disk.  If the inclination measured via the blurred reflection component is correct, our line of sight may align with the launching angle of the warm absorbers.  Then, the contrast between the expected velocity ($-v = 870~{\rm km}~{\rm s}^{-1})$ and observed absorption velocity ($-v = 290~{\rm km}~{\rm s}^{-1}$) may reveal more about the flow.  Circulating gas will tend to spend most of its time close to apocenter, where its velocity is low.  Indeed, for the implied radii and velocities, the turnaround radius is $r_{turn}/r_{launch} \lesssim 2$.  It is possible, then, that warm absorbers are circulating flows that are typically observed close to their turnaround point.

\subsection{The Fe K edge}

Our best-fit model for the Fe~K edge requires cool gas, with a temperature of $kT = 5.5^{+0.4}_{-0.3}$~eV or $T = 6.4^{+0.5}_{-0.4}\times 10^{4}$~K.  The charge states present in this gas have edges at slightly higher energies than neutral Fe (7.112~keV), and provide a far better match to the summed Resolve spectrum.  Figure 5 shows the contrast between our best fit model, and one that attempts to describe the edge through cold, entirely neutral gas. The best fit with a neutral edge gives a fit statistic of $C = 5391$, or $\Delta C = 191$ relative to the model in Table 2.  The difference gives $\Delta{\rm AIC} = -189$.  The measured gas temperature exceeds typical values for the optical broad line region by a factor of a few, and we examine this more in the next section.

Following \cite{miller2025b}, we searched for evidence of dust in the Fe~K edge using the ``amol'' model within SPEX, which includes cross sections for numerous abundant minerals and compounds.  The most promising of these is likely olivine: it is detected in interstellar dust, and it is the most common mineral in Earth's mantle.  Its cross section introduces a sharp, line-like trough to the Fe~K edge close to the threshold energy \citep{rogantini2018}.  It is this deviation from a step function that makes a clear test possible.  To our knowledge, no other common compound offers an equally distinctive edge structure.

The most compelling residuals in the Fe~K edge may be a pair of line-like features at 7.2~keV (see Figure 5).  In order to describe these features in terms of olivine, the dust would have to be blue-shifted by $-v=2900~{\rm km}~{\rm s}^{-1}$.  This would rank as the fastest molecular flow yet detected in a Seyfert or quasar, and would represent an implausible mass flux.  Not only do the data not statistically require the addition of olivine to the local model, but the columns required to match even these weak features predicts low-energy attenuation that is readily rejected by the observed continuum.

Dust may contribute to the production of Fe~K$_{\alpha}$ emission lines at the base and/or inner faces of the BLR and torus, but is not clear that it should be retained in the more diffuse and highly ionized absorbing gas above the disk.  Self-consistent treatments of dust include low-energy absorption, and even modest columns strongly attenuate low-energy spectra.  Even at the sensitivity afforded in this summed 0.9~Ms spectrum of NGC 4151, there is no clear evidence of dust absorption in the Fe~K edge structure.  

\subsection{Baked Alaska?}

The model detailed in Table 2 assumes that the partial covering column density belongs to a gas zone that lies outside of the central engine, covering the continuum, Fe~K emission line regions, and wind regions.  At $kT = 5.5^{+0.4}_{-0.3}$~eV, it is cooler than even the least ionized warm absorber ($kT = 0.23$~keV for ``WA-2''), but several times hotter than temperatures associated with the optical BLR.  Moreover, this gas is likely a driver of the state-like variability that is observed in NGC~4151 (see, e.g, Figure 1 in \citealt{Miller_2018}).  The time scale of that variability is merely several weeks.  Taken together, it is at least possible that the partial covering obscuration lies between the inner disk and BLR, potentially representing a disk atmosphere that responds to the central engine and modulates the X-ray flux that is transmitted to the BLR, torus, and winds.  

To test whether this is consistent with the 0.9~Ms Resolve spectrum of NGC~4151, we simply reordered the absorbing layers within SPEX and placed the ``hot'' component (modeling partial covering absorption) closest to the central engine.  In this scheme, the Fe~K emission line and wind regions only see flux that has passed through this zone.  The cool disk and cool absorber are central, while the hot winds are more distant, similar to a baked Alaska dessert wherein ice cream is surrounded by a hot meringue.  This model achieves a Cash statistic of $C = 5206$ for 4858 degrees of freedom, representing an equally good fit to the data.  

If the partial covering absorber is as close to the central engine as the ionized winds, it may be photoionized.  To test if this possibility is also consistent with the data, we replaced the inner ``hot'' component with an additional ``pion'' component.  An excellent fit is achieved, giving a Cash statistic of $C = 5200.0$ for 4856 degrees of freedom.  We measure of a column density of $N_{H} = 1.2\pm 0.1\times 10^{23}~{\rm cm}^{3}$, an ionization parameter of ${\rm log}\xi = -1.9^{-0.08}_{+0.5}$, a covering factor of $f_{c} = 0.86\pm 0.01$, and a bulk velocity shift of $v = 0^{+2800}_{-200}~{\rm km}~{\rm s}^{-1}$.  Here, zero shift corresponds to an ionization parameter of ${\rm log}\xi = -1.9$, while an ionization parameter of ${\rm log}\xi = 1.4$ requires a red-shift of $v = 2800~{\rm km}~{\rm s}^{-1}$.  The lower ionization parameter bulk velocity shift are likely more physically plausible.

Whether the absorption is photoionized or collisional, the fact of low-ionization absorption rather than neutral absorption predicts absorption lines and edges at low luminosity. These are not evident within the Resolve spectrum, likely owing to the multiple layers of absorption along our line of sight to the central engine.  At lower energies, the X-ray spectrum of NGC~4151 is dominated by emission lines from the prominent narrow-line region (\citealt{kraemer2005},\citealt{wang2011}).

While the edge structure itself is inconsistent with neutral gas and does not reveal clear contributions from dust, it is important to note some key systematic uncertainties.  \cite{kallman2004} show that ionized Fe~K edges are not sharp step functions; rather, they are shaped by a series of resonance features from $1s-np$ autoionizing transitions.  The effects are largely negligible at ${\rm log}\xi = -2$, similar to what our fits measure.  However, two strong resonance features separated by $\leq 20$~eV are predicted for ${\rm log}\xi = -1.25$.  Conservatively, a 20~eV smearing would translate to an additional velocity uncertainty of $\Delta v \leq 800~{\rm km}~{\rm s}^{-1}$.  If the gas is collisional, this level of smearing would allow for a broader range of gas temperatures, representing a systematic error of $\Delta E = \pm1.5$~eV or $\Delta {\rm T} = \pm 1.75\times 10^{4}$~K.

The radii of the BLR and torus regions probed by the Fe~K$_{\alpha}$ emission line components and the velocities of the wind components are unaffected by positioning the absorber close to the central engine.  The ionization parameters of the wind components all drop substantially as the transmitted X-ray continuum is lower when the model is constructed in this way.  Whether the emission line regions see the central engine directly or only after passing through a partial covering zone, might be determined by a careful variability and time lag analysis; this is deferred to a subsequent analysis.

\section{Discussion}
\label{sec:discussion}

We have analyzed the time-averaged Resolve spectrum of NGC 4151, comprised of five observations during the PV phase and nine segments from Cycle 1.  The resulting 0.9~Ms spectrum is likely to be among the most sensitive AGN spectra that XRISM will obtain during its operations, and it holds some important clues for how spectra from NGC~4151 and Seyferts can be modeled and utilized.  In this section, we examine how this analysis has advanced our understanding of the accretion flow structure and feedback through accretion-driven winds.  We also examine some of the limitations of our analysis, look at outstanding questions that remain, and comment on the logical next steps for NGC 4151 and other Seyferts in the XRISM era.

\subsection{Emission from the Inner Disk?}

Our models suggest that relativistic reflection from the innermost disk may be visible in NGC~4151.  Potentially, the need for this emission has been clarified at high spectral resolution via the sensitivity of XRISM to narrow emission and absorption lines, relative to prior CCD spectra.  Alternatively, the curvature the we have modeled in terms of relativistic reflection might be an artifact of summing many spectra that differ slightly (in power-law index, flux, and internal partial covering column density and covering factor), or a misinterpretation of scattered continuum emission.  If the reflection signatures are robust, contradictions with lag spectra would signal that the latter are compromised by scattering.

Lag spectra have been utilized with the goal of finding a model--independent support for relativistic reflection from the inner disk (e.g., \citealt{zoghbi2010}, \citealt{kara2016}).  Some Seyferts that show putative relativistic lines in their time-averaged spectra -- such as MCG-6-30-15 \citep{kara2014} -- do not show relativistic lines in their lag spectra.  NGC~4151 emerged as the rare (or, singular) case with evidence of a relativistic line in both time-averaged and lag spectra (e.g. \citealt{zoghbi2012}), and permitted the extraction of a relativistic transfer function \citep{cackett2014}.  However, recent work demonstrates that the lag spectra of NGC 4151 are consistent with continuum scattering, and do not require relativistic reflection \citep{Zoghbi_2019}.  

Even more vividly than the 0.9~Ms spectrum treated in this work, individual XRISM observations of NGC~4151 \cite{xiang2025} require a layered and variable line of sight through the central engine.  The data even suggest {\em axial asymmetry} in the wind structure.  Relativistic reflection from the inner disk must filter through this complex wind environment, incurring a number of delays (and, the delays may also vary).  Within this environment, it may be natural for the {\em time-averaged} spectra to reveal relativistic reflection, while the {\em lag} spectra would not.  Evidence of a broad, potentially relativistic lines in the Resolve spectra of NGC~3783 and MCG-6-30-15 may support this idea (\cite{mehdipour2025}, \citealt{li2025}, \citealt{brenneman2025}). 

\subsection{The broad line region and torus}

Utilizing self-consistent Fe~K$_{\alpha}$ and K$_{\beta}$ line kernels that include scattering from bound electrons, we are able to model the narrow emission line complex in NGC~4151 only with contributions from the torus and optical BLR.  The smooth red shoulder on the line cores is modeled without invoking a structure at $r\simeq 100~GM/c^{2}$, in contrast to \cite{2024ApJ...973L..25X}.  It is possible that the new line kernels are incorrect and that a warp and/or wind-launching (UFO-launching?) region {\em does} exist at such a small radius, but it is an extreme possibility that is more simply explained through atomic effects.    

The statistical preference for an emissivity index of $q=2$ over $q=3$ indicates that -- at least with the improved line kernels -- the data are sensitive to the vertical extent of the BLR and torus above the midplane of the system.  This is consistent with the ``bowl'' geometry that is inferred in some optical and IR studies (e.g., \citealt{gravity2024}).    

As shown in Table 2, the BLR and torus line components have inner radii of $r_{BLR} = 1.9^{+1.3}_{-0.5}\times 10^{3}~GM/c^{2}$ and $r_{tor} = 2.8^{+2.0}_{-1.0}\times 10^{3}~GM/c^{2}$.  Despite the different line kernels and best-fit emissivities in this work and \cite{2024ApJ...973L..25X}, these radii are consistent with the prior XRISM measurements.  The inner radius of the BLR reported in Table 2 is a factor of $\sim2$ smaller than the radii inferred via direct fits to broad UV emission lines \citep{bentz2015}, potentially indicating an inner X-ray extent to the UV and optical BLR.  The inner radius measured for the torus is formally consistent with dust reverberation mapping and IR interferometry of NGC~4151 (\citealt{lyu2021}, \citealt{gravity2023}). 

Nevertheless, the fact that the data require only a modest separation of the BLR and torus may signal that the two are not fully distinct, but rather connected.  For instance, the BLR and torus may mark the transition between a gas-dominated disk that broadly follows the predictions of \cite{ss73} to a dusty disk with larger scale height.  Similar transition regions are well known in disks around T Tauri stars, where curved interfaces with varying compositions can be inferred (e.g., \citealt{espaillat2010}, \citealt{mcclure2013}).  In this case, the BLR may mark the point where dust starts to become important at low scale heights but cannot survive when exposed to more direct radiation at larger heights, whereas the torus may mark the point at which dust is important at all scale heights.  Prominent models suggest that dust may be important to raising gas above the plane of the disk in the BLR (see, e.g., \citealt{czerny2015}), and our results appear to support this picture.  The data may also support models that account for AGN obscuration  through clumpy wind zones.  \cite{xiang2025} find multiple lines of evidence suggesting that the winds in NGC~4151 are clumpy and potentially driven magnetically;   \cite{elitzur2006} describe a scenario wherein clumpy magnetohydrodynamic winds give rise to the geometry we call the torus.

\subsection{Variable, cool, partial-covering absorption}

In the 0.9~Ms Resolve spectrum of NGC~4151, the Fe~K edge does not lie at the energy expected for neutral Fe (7.112~keV), and it is not sharp (see Figure 5).  Rather, the edge structure can be characterized in terms of cool, $kT = 5.5^{+0.4}_{-0.3}$~eV gas that exhibits a set of absorption edges from several charge states (but note that it is equally well described in terms of photoionized gas).  Does internal obscuration within Seyferts generally have this character?  

Even with Resolve, there are very few sources wherein the Fe~K edge can be studied in detail.  Mrk~279 is a Seyfert-1.5 AGN with emission line properties similar to those found in NGC~4151, but it is much fainter \citep{miller2025}.  The Resolve spectrum of NGC~3783 shows a slope close to the Fe~K edge, rather than a sharp feature; this is interpreted as part of a relativistic emission line \citep{li2025}, but part of this feature could be due to cool partial covering absorption.  The best point of comparison might be NGC~3516, a Seyfert-1.5 AGN with a variable internal column similar to that observed in NGC~4151.  The Resolve spectrum of NGC~3516 also appears to require a smooth Fe~K edge structure rather than a sharp, neutral edge \citep{juranova2025}.  For all of these features to be relativistic lines, the inclinations and spins of different AGN would need to always locate blue wings close to the neutral edge; on balance, it may be simpler if these features arise through atomic processes.

Our model achieves a good fit to the spectra, assuming that the cool, partially covering column density is radially exterior to the ionized absorption zones (see Table 2).  This assumption may be inconsistent with variations in this absorber, which play a role in determining the overall ``state'' of NGC~4151.  Fits that instead place this zone closest to the innermost disk -- interior to even the UFOs -- achieve an equally good Cash statistic (see Section 4.7).  Close to the black hole, the gas is more likely to be photoionized, and this is also entirely consistent with the data.  It is therefore possible that the cool variable absorber lies interior to some or all of the wind components and/or narrow Fe~K$_{\alpha}$ line production regions, and that the emission lines (in particular) see an attenuated flux.  This would affect how we perceive their response to changes in the ionizing flux from the central engine.  

The edge structure cannot be fit by replacing the torus and BLR line kernels with full reflection models.  The broadening required to fit emission from the BLR is insufficient to match the edge profile.  Such models also fail to fit the red wing of the narrow Fe~K$_{\alpha}$ line structure, as they anticipate a sharp shoulder at 6.25~keV. Ionized reflection models cannot simultaneously account for the narrow emission line components and edge, because the spacing of the Fe~K$_{\alpha}$ and K$_{\beta}$ lines signals neutral gas, whereas the edge requires ionization structure.

\subsection{Dust}

Among known, common compounds, olivine likely presents the best chance for direct detection via absorption close to the Fe K edge.  However, the data appear to strongly reject its inclusion within the model.  This is not the same as rejecting the presence of dust within the accretion flow.  Indeed, the smooth scattering shoulder on the narrow Fe~K$_{\alpha}$ emission line complex may represent indirect evidence of dust.  Moreover, it is the presence of dust at the innermost face of the torus that enables IR reverberation mapping in AGN (e.g., \citealt{minezaki2019}).  

Indications of dust in the emission line structure and apparent absence in absorption may simply reflect a situation wherein dust is present close to the plane of the disk in the BLR, where natal winds are taking shape (e.g., \citealt{czerny2015}) -- and at the face of the torus -- but {\em not} at larger scale heights.  Strong mass-loading of the X-ray winds with dust might prevent them from reaching the velocities necessary to escape the central engine, and therefore might also be incompatible with the presence of X-ray VFOs and UFOs in NGC 4151 and other AGN.  The presence of dust at larger scale heights might also preclude the detection of broad UV lines.

\subsection{UFOs}

The UFOs that are observed to vary in the five PV observations \citep{xiang2025} are evident in the summed 0.9~Ms spectrum (see Table 2, and Figures 1, 2, and 5).  The wind properties seen in the summed spectrum may represent average values for these flows; in this respect, they may be a good guide to the level of feedback that these flows deliver to the host galaxy over long periods.  

For volume filling factors of unity, the mass outflow rate and kinetic power in UFO-1 greatly exceed the mass accretion rate and radiative luminosity of NGC~4151, respectively.  For the lower limit of $f_{v} = 5\times 10^{-4}$ estimated by \cite{xiang2025}, UFO-1 falls short of producing the threshold to halt star formation in the host bulge, $L_{kin} = 0.5-5\% L_{Edd}$ (\citealt{dimatteo2005}, \citealt{hopkins2010}).  This is likely because we have measured a higher ionization parameter in UFO-1 than \cite{xiang2025} measured in this wind component, reducing the kinetic power by $\xi^{-1}$.  A conservative interpretation of these outcomes is that UFO-1 does not clearly represent galaxy-altering feedback on average, but can clear the threshold if the filling factor is not minimial and/or if the wind properties vary with time.

UFO-2 is much less energetic, and less likely to deliver strong feedback.  In Figure 5, it is clear that UFO-2 absorbs some of the narrow Fe~K$_{\beta}$ emission line flux (also see \citealt{xiang2025}).  On this basis, it is tempting to conclude that the UFO may be observed at a point that is radially exterior to the narrow emission line zones.  However, this conclusion may be flawed, in part or in whole.  {\em Emission lines} likely originate from the full $2\pi$ cylinder at a given radius, so more emission line flux originates from the far side of the central engine than along the pencil-beam that accumulates absorption on the near side.  Depending on the geometry of the accretion inflow and connected winds, absorption close to the black hole but on the near side of the central engine can diminish emission lines generated farther from the black hole on the far side.

\subsection{Warm Absorbers}

The 0.9~Ms Resolve spectrum has enabled significant progress on the nature of warm absorbers in Seyfert AGN.  The sensitivity of the spectrum reveals an emission component to the most ionized warm absorber (WA-1 in Table 2).  The broadening ($970^{+160}_{-90}~{\rm km}~{\rm s}^{-1}$) and velocity shift ($\sigma = 870\pm 30~{\rm km}~{\rm s}^{-1}$) of this component determine its radius ($r = 9.6^{+2.0}_{-2.5}\times 10^{4}~GM/c^{2}$), density ($n = 6.8^{+5.6}_{-2.2}\times 10^{4}~{\rm cm}^{-3}$), and filling factor ($f \simeq 0.6$) through the ionization parameter formalism.  Comparing the velocity shifts of this warm absorber to local escape velocities, and comparing the small velocity shift seen in absorption to the larger shift seen in emission, the most ionized warm absorber is likely a failed wind that is observed close to apocenter.  The large measured covering factor of this zone ($\Omega/4\pi = 1.0_{-0.1}$), and its radial location -- exterior to even the innermost face of the torus -- may explain why warm absorbers are commonly observed in Seyfert-1 AGN and even some Seyfert-2 cases (e.g., NGC 4388; see \citealt{miller2019}).  

A second warm absorber (WA-2 in Table 2) is observed with very similar velocities, suggesting a multi-phase gas.  The more highly ionized warm absorber may pressure-confine the lower-ionization warm absorber, and fill more of the local volume.  This may explain the lack of a clear emission component in the lower-ionization absorber.  We note that the photoionization model for WA-2 gives a temperature of $kT = 0.23$~keV, or $T = 2.7\times 10^{6}$~K.  This may preclude the presence of dust in the warm absorbers, though dust may be intercepted at other points along the line of sight.

\subsection{Future directions}

We have applied physically motivated models to measure properties of the emission regions and outflows in NGC~4151.  However, the sensitivity afforded by 0.9~Ms of exposure on the brightest Seyfert-1 AGN in the XRISM pass band is extraordinary, and it is not clear that our models have harnessed all of the information within the spectrum.  
An improved understanding of the emission regions may require combining the framework that underlies models such as XSTAR with a tunable geometrical blueprint.  

A publicly available emission model that explicitly includes key geometric quantities as fitting parameters (e.g., BLR and torus inner radius, opening angle, solid angle) may offer more insights than disconnected components and quantities such as the emissivity index.  Rather than a single column density, it may be helpful to constrain parameters that control a run of column density with radius and height above the midplane.  Also allowing for gradients in ionization parameter may even make it possible to constrain the geometric parameters by simultaneously considering optical and UV emission lines.

Similarly, geometric wind models may now be in order.  For instance, \cite{xiang2025} found that the different wind components in NGC 4151 may all be consistent with a magnetocentrifugal wind.  The properties of these winds are largely determined by the conditions at the point where they are launched (see, e.g., \citealt{fukumura2010}).  A publicly available model that allows the data to constrain wind launching radii, velocities, and opening angles in terms of gas properties may enable enormous progress.  

The winds observed in NGC~4151 with XRISM are more highly ionized than those traced with Chandra.  \cite{kraemer2005} report the detection of seven absorption components in HETG spectra, and \cite{couto2016} study variations in these components.  Assuming a typical AGN SED, ${\rm log}\xi = {\rm log}U + 1.7$ \citep{chakraborty2021}, these absorbers then span $0.03 \leq {\rm log}\xi \leq 2.75$.  However, based on their properties and variability, \cite{kraemer2005} and \cite{couto2016} suggest that some wind components may be driven magnetically.  \cite{behar2009} did not include NGC~4151 in the small set of AMDs collected from Seyerts based on Chandra data.  Future work that examines the AMD of prior Chandra data from NGC~4151, separately and jointly with XRISM data, may be able to develop a more complete picture of wind driving in this AGN.

Finally, it is not yet clear if the UFOs in NGC~4151 represent galaxy-altering feedback. For the most conservative volume filling factors, these wind components may only supply the needed kinetic power in specific intervals.  However, the UFOs may typically deliver galaxy-altering feedback into the host bulge if the volume filling factor takes moderate values. Future work must focus on the duty cycle of UFOs in NGC~4151 and other AGN, and on measuring the number density and filling factor of gas in these critical flows.

\section{Conclusions}

\noindent$\bullet$ The narrow Fe~K$\alpha$ line structure in NGC~4151 can be described in terms of neutral gas located at the irradiated inner faces of the BLR and torus. 

\noindent$\bullet$ The smooth red wing on the line cores from these regions is successfully modeled in terms of Compton scattering from electrons bound in cold gas and/or dust.  This may serve as an indirect detection of dust in the BLR, consistent with models that rely on radiation pressure to lift BLR gas above the plane of the disk (e.g., \citealt{czerny2015}).  These results may obviate models invoking a warp and/or wind originating from $r\simeq 100~GM/c^{2}$ to describe the smooth red wing.

\noindent$\bullet$ An isotropic, point-source central engine irradiating a truly flat disk in the equatorial plane gives rise to an emissivity index of $q = 3$ (where $J \propto r^{-q}$). 
The BLR and torus components of the Fe~K emission line prefer an emissivity of $q=2$.  This signals that the data are sensitive to the vertical extent of these geometries above the equatorial plane.   
The overall geometry indicated by the emission line structure is likely then consistent with a ``bowl,'' similar to recent results gleaned in optical and IR bands.

\noindent$\bullet$ The Fe~K region and high-energy portion of the Resolve spectrum are well fit when relativistic, ionized reflection from the innermost accretion disk is included.  At least within the framework of our models, the Resolve data alone are not able to constrain the spin parameter of the black hole.

\noindent$\bullet$ The relativistic reflection component may not be a unique description of broad curvature within the spectrum.  Especially in view of lag studies that find no evidence of relativistic reflection in NGC~4151 -- but evidence of scattered light -- some of the observed curvature may be due to interactions in the complex set of wind zones that light from the central engine encounters before it is detected.

\noindent$\bullet$ The Fe~K edge in NGC~4151 is likely composed of features from multiple charge states in cool gas, rather than a single step-function edge from cold, neutral gas.  Given that this gas column varies over weeks and helps to determine the broad X-ray character of NGC 4151, it is natural that it may be situated within the central engine and elevated in temperature.

\noindent$\bullet$ The Fe~K edge structure in NGC 4151 does not require contributions from dust.  Modeling weak features close to 7.2~keV in terms of olivine, for instance, does not yield statistical improvements, and would require the fastest molecular outflow yet detected in an AGN.  

\noindent$\bullet$ Slow, $-v \simeq 200~{\rm km}~{\rm s}^{-1}$ absorption zones in NGC~4151 almost certainly represent failed winds that circulate within the central engine and do not escape to infinity (also see \citealt{xiang2025}).

\noindent$\bullet$ The most highly ionized of the slow wind zones is primarily seen in Fe~XXVI, and includes a clear emission component that subtends a solid angle consistent with $\Omega/4\pi = 1$.  If the observed broadening of the emission component is due to Keplerian motions, the gas is located at distance similar to the torus.  This distance is commensurate with radii inferred via the ionization parameter, suggesting a high volume filling factor, and only limited clumping in this component.

\noindent$\bullet$ The motions of failed winds are slowest at apocenter, where such flows stall. Slow X-ray ``warm absorbers'' may typically be observed close to their turn-around point.

\noindent$\bullet$ The summed 0.9~Ms Resolve spectrum reveals clear evidence of very fast outflows, and ultra-fast outflows.  However, these zones modify the summed spectrum less than in individual spectra, because the fast wind zones are highly variable \citep{xiang2025}.  

\noindent$\bullet$ For the most conservative volume filling factors, even the fastest UFO in NGC~4151 may not deliver the kinetic power needed to halt star formation in the host bulge.  However, this UFO may occasionally represent galaxy-altering feedback, and would regularly represent strong feedback for moderate volume filling factors.

\hspace{0.2in}
We thank the anonymous referee for comments that improved this manuscript.  JMM acknowledges helpful conversations with Niel Brandt and Michael Eracleous, and SPEX help from Jelle de Plaa.

\bibliography{main}{}
\bibliographystyle{aasjournalv7}

\begin{deluxetable}{lll}
    \tablenum{1}
    \tablecaption{\label{table:observations}}
    \tablehead{\colhead{ObsID} &\colhead{Start Time}&\colhead{Net Exposure}}
    \startdata
       ~  & (MJD) & (ks) \\
    \hline   
    000125000 & 60280.3 & 72 \\
    000137000 & 60304.8 & 56 \\
    300047020 & 60448.3 & 98 \\
    300047030 & 60476.9 & 83 \\
    300047040 & 60483.6 & 89 \\
    \hline
    201076010 & 60656.9 & 55 \\
    201076020 & 60659.0 & 56 \\
    201076030 & 60661.2 & 69 \\
    201076040 & 60663.5 & 70 \\
    201076050 & 60665.8 & 60 \\
    201076060 & 60668.0 & 52 \\
    201076070 & 60670.5 & 47 \\
    201076080 & 60672.3 & 42 \\
    201076090 & 60674.1 & 44 \\
    \hline
    total     &   --    & 893 \\
    \hline
    \enddata
\end{deluxetable}

\begin{figure*}
    \centering
    \includegraphics[width=1.0\textwidth]{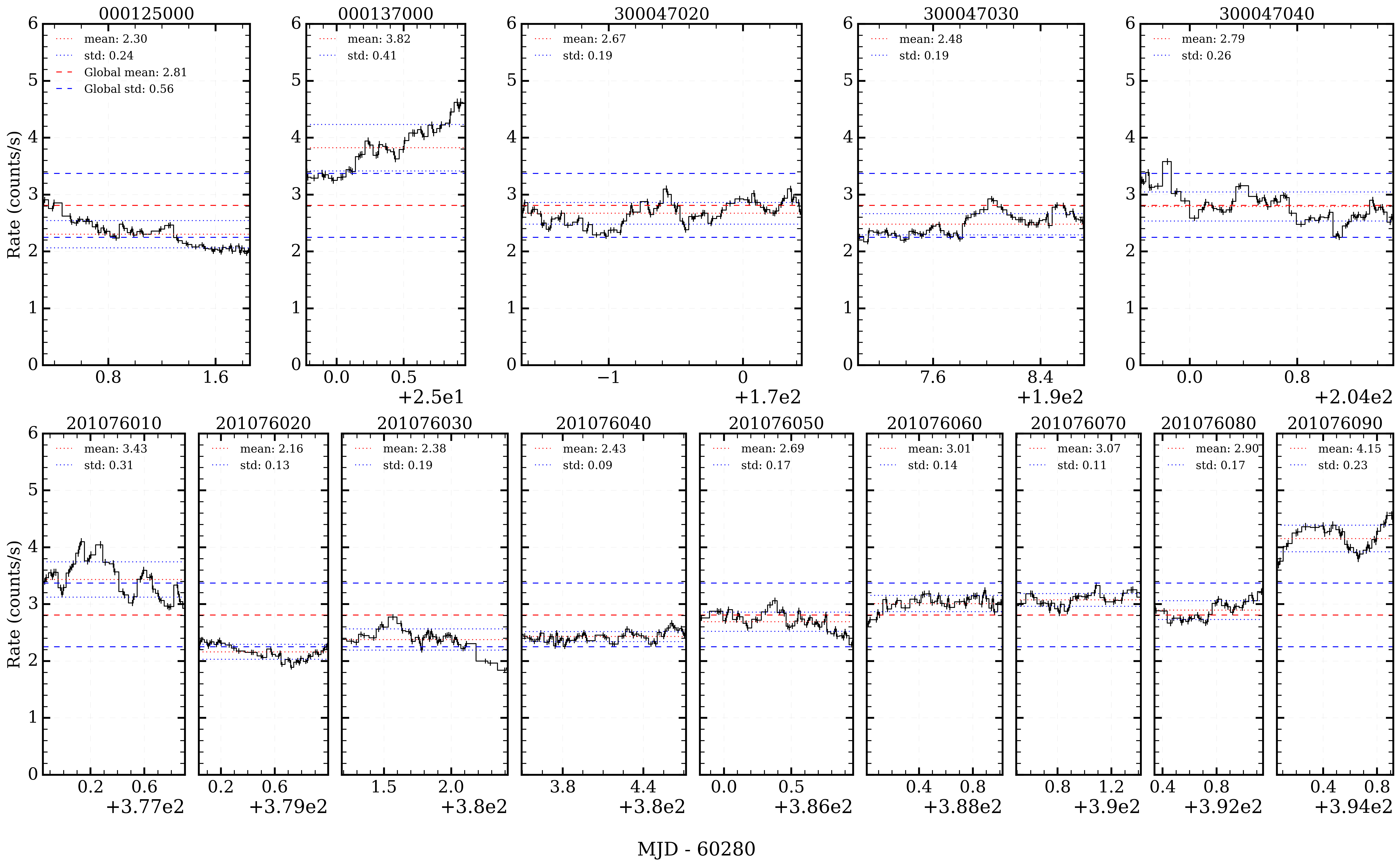}
    \caption{Light curves of the 14 Resolve exposures of NGC~4151 considered in this work.  Each light curve covers the 2.2--17.4~keV band and the time bins are 1024~seconds.  The first five exposures were obtained randomly but the final nine observations were obtained ever other day for 18 days.}
    \label{fig:lightcurves}
\end{figure*}

\clearpage

\begin{table}[t!]
\caption{Key Spectral Model Parameters}
\begin{scriptsize}
\begin{center}
\begin{tabular}{lll}
Parameter & Value & Comments\\
\tableline    
{\it X-ray continuum} & -- & -- \\
power-law $\Gamma$ & $1.510^{+0.002}_{-0.002}$ & -- \\
power-law norm. ($10^{51}~{\rm ph}~{\rm s}^{-1}~{\rm keV}^{-1}$) &  $1.197^{+0.001}_{-0.001}$ & norm. at 1~keV \\
\tableline
{\it Inner disk line component} & -- & speith~$\times$~xillverHR \\
$r_{in}$ ($GM/c^{2}$) & $3.2^{+3.5}_{-2.0}$ & -- \\
$i$ (deg.) & $29.7^{+0.5}_{-0.4}$ & --\\
$q$ & $3.0^{*}$ & fixed \\
${\rm log}\xi$ & $3.00^{+0.01}_{-0.15}$ & -- \\
Norm. & $192.5^{+6.9}_{-9.6}$ & -- \\
\tableline
{\it BLR line component} & -- & speith~$\times$~xclumpy \\
$r_{in}$ ($10^{3}~GM/c^{2}$) & $1.9^{+1.3}_{-0.5}$ & -- \\
$i$ (deg.) & $48^{+38}_{-9}$ & -- \\
$q$ & $2.0^{*}$ &  fixed \\
${\rm N}_{\rm H}$ ($10^{24}~{\rm cm}^{-2}$) & $1.4^{+7.6}_{-0.3}$ & -- \\
Norm. ($10^{6}~{\rm ph}~{\rm cm}^{-2}~{\rm s}^{-1}~{\rm keV}^{-1}$) & $9.9^{+3.7}_{-0.1}$ & -- \\
\tableline
{\it torus line component} & -- & speith~$\times$~xclumpy \\
$r_{in}$ ($10^{3}~GM/c^{2}$) & $2.8^{+2.0}_{-1.0}$ & -- \\
$i$ (deg.) & $9^{+1}_{-1}$ & -- \\
$q$ & $2.0^{*}$ & fixed \\
${\rm N}_{\rm H}$ ($10^{24}~{\rm cm}^{-2}$) & $0.4^{+0.5}_{-0.2}$ & -- \\
Norm. ($10^{6}~{\rm ph}~{\rm cm}^{-2}~{\rm s}^{-1}~{\rm keV}^{-1}$) & $4.5^{+3.9}_{-0.9}$ & -- \\
\tableline
{\it UFO-1} & -- & pion \\
${\rm N}_{\rm H}$ ($10^{24}~{\rm cm}^{-2}$) & $0.022^{+0.011}_{-0.005}$ & -- \\
${\rm log}\xi$ & $4.00^{+0.15}_{-0.09}$ & -- \\
$\sigma$ ($10^{3}~{\rm km}~{\rm s}^{-1}$) & $4.4^{1.3}_{-0.9}$ & -- \\ 
$-v$ ($10^{3}~{\rm km}~{\rm s}^{-1}$) & $50.3^{+1.2}_{-1.2}$ & -- \\
\tableline
{\it UFO-2} & -- & pion \\
${\rm N}_{\rm H}$ ($10^{24}~{\rm cm}^{-2}$) & $0.0011^{+0.0003}_{-0.0003}$ & -- \\
${\rm log}\xi$ & $3.5^{+0.1}_{-0.1}$ & -- \\
$\sigma$ ($10^{3}~{\rm km}~{\rm s}^{-1}$) & $0.16^{+0.6}_{-0.4}$ & -- \\ 
$-v$ ($10^{3}~{\rm km}~{\rm s}^{-1}$) & $16^{+1}_{-1}$ & -- \\
\tableline
{\it VFO-1} & -- & pion \\
${\rm N}_{\rm H}$ ($10^{24}~{\rm cm}^{-2}$) & $0.0063^{+0.0008}_{-0.0008}$ & -- \\
${\rm log}\xi$ & $3.21^{+0.01}_{-0.01}$ & -- \\
$\sigma$ ($10^{3}~{\rm km}~{\rm s}^{-1}$) & $1.1^{+0.5}_{-0.2}$ & -- \\ 
$-v$ ($10^{3}~{\rm km}~{\rm s}^{-1}$) & $3.6^{+0.2}_{-0.4}$ & -- \\
\tableline
{\it WA-1} & -- & pion \\
${\rm N}_{\rm H}$ ($10^{24}~{\rm cm}^{-2}$) & $0.020^{+0.001}_{-0.002}$ & -- \\
${\rm log}\xi$ & $3.75^{+0.03}_{-0.03}$ & -- \\
$\sigma$ ($10^{3}~{\rm km}~{\rm s}^{-1}$) & $0.22^{+0.01}_{-0.01}$ & -- \\ 
$-v$ ($10^{3}~{\rm km}~{\rm s}^{-1}$) & $0.29^{+0.01}_{-0.02}$ & -- \\
$\Omega/4\pi$ & $1.0^{*}_{-0.1}$ & emission covering factor \\
$\sigma$ ($10^{3}~{\rm km}~{\rm s}^{-1}$) & $0.97^{+0.16}_{-0.09}$ & emission broadening \\
$v$ ($10^{3}~{\rm km}~{\rm s}^{-1}$) & $0.87^{+0.03}_{-0.03}$ & emission shift \\
\tableline
{it WA-2} & -- & pion \\
${\rm N}_{\rm H}$ ($10^{24}~{\rm cm}^{-2}$) & $0.0072^{+0.0006}_{-0.0004}$ & -- \\
${\rm log}\xi$ & $3.259^{+0.007}_{-0.006}$ & -- \\
$\sigma$ ($10^{3}~{\rm km}~{\rm s}^{-1}$) & $0.19^{+0.02}_{-0.01}$ & -- \\ 
$-v$ ($10^{3}~{\rm km}~{\rm s}^{-1}$) & $0.27^{+0.02}_{-0.01}$ & -- \\
\tableline
{\it partial covering absorption} & -- & via the ``hot'' model \\
${\rm N}_{\rm H}$ ($10^{24}~{\rm cm}^{-2}$) & $0.118^{+0.001}_{-0.001}$ & -- \\
kT (eV) & $5.5^{+0.4}_{-0.3}$ & -- \\
${\rm f}_{\rm cov}$ & $0.853^{+0.002}_{-0.002}$ & geometric covering factor \\
\tableline
Cash Statistic / degrees of freedom & 5200/4848 & -- \\
\tableline
\end{tabular}
\end{center} 
\tablecomments{Key spectral parameters and errors for the best-fit model of the 0.9~Ms Resolve spectrum of NGC~4151.  The order in which the components are listed reflects their physical role: X-rays from the central engine excite emission lines, and the continuum and emission lines then pass through a sequence of layered wind zones and stationary obscuration.  Proto-solar abundances were assumed in all components \citep{lodders2009}.  In the ``pion'' components, an absorption covering factor of ${\rm f}_{\rm cov} = 1$ was adopted.  ``Warm absorber flow 1'' includes re-emission from gas with the same column, ionization, and intrinsic broadening, at zero velocity shift.  An UV blackbody component based on fits to Hubble/STIS data is included in the total model and feeds into the photoionization models.  The total ionizing luminosity is $L_{ion} = 9.2\times 10^{43}~{\rm erg}~{\rm s}^{-1}$.  See the text for additional details.} 
\end{scriptsize}
\end{table}

\clearpage

\begin{figure*}
    \centering
    \includegraphics[width=1.0\textwidth]{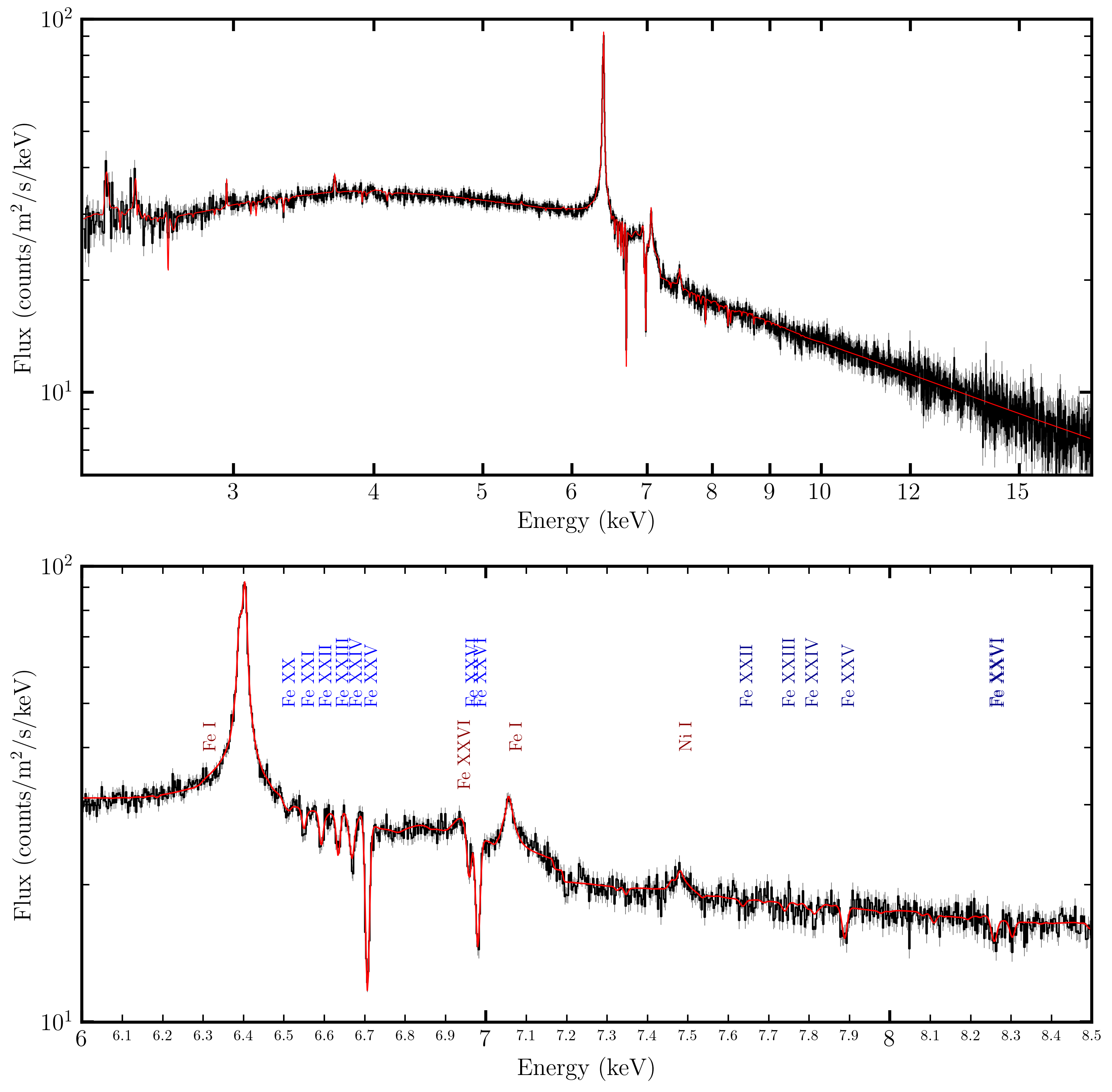}
    \caption{Top: The 0.9~Ms Resolve spectrum of NGC 4151 over the full fitted pass band.  Bottom: The spectrum in the Fe~K band.  The data and models are plotted in the host frame.  The model in red is detailed in Table 2.  The key components include Fe~K emission line flux associated with the torus, broad-line region, and potentially the inner disk, and several zones of ionized absorption including ultra-fast outflows.  Prominent narrow emission and absorption lines are labeled.}
    \label{fig:broadfek}
    \vspace{-0.5in}
\end{figure*}

\clearpage

\begin{figure*}
    \centering
    \includegraphics[width=1.0\textwidth]{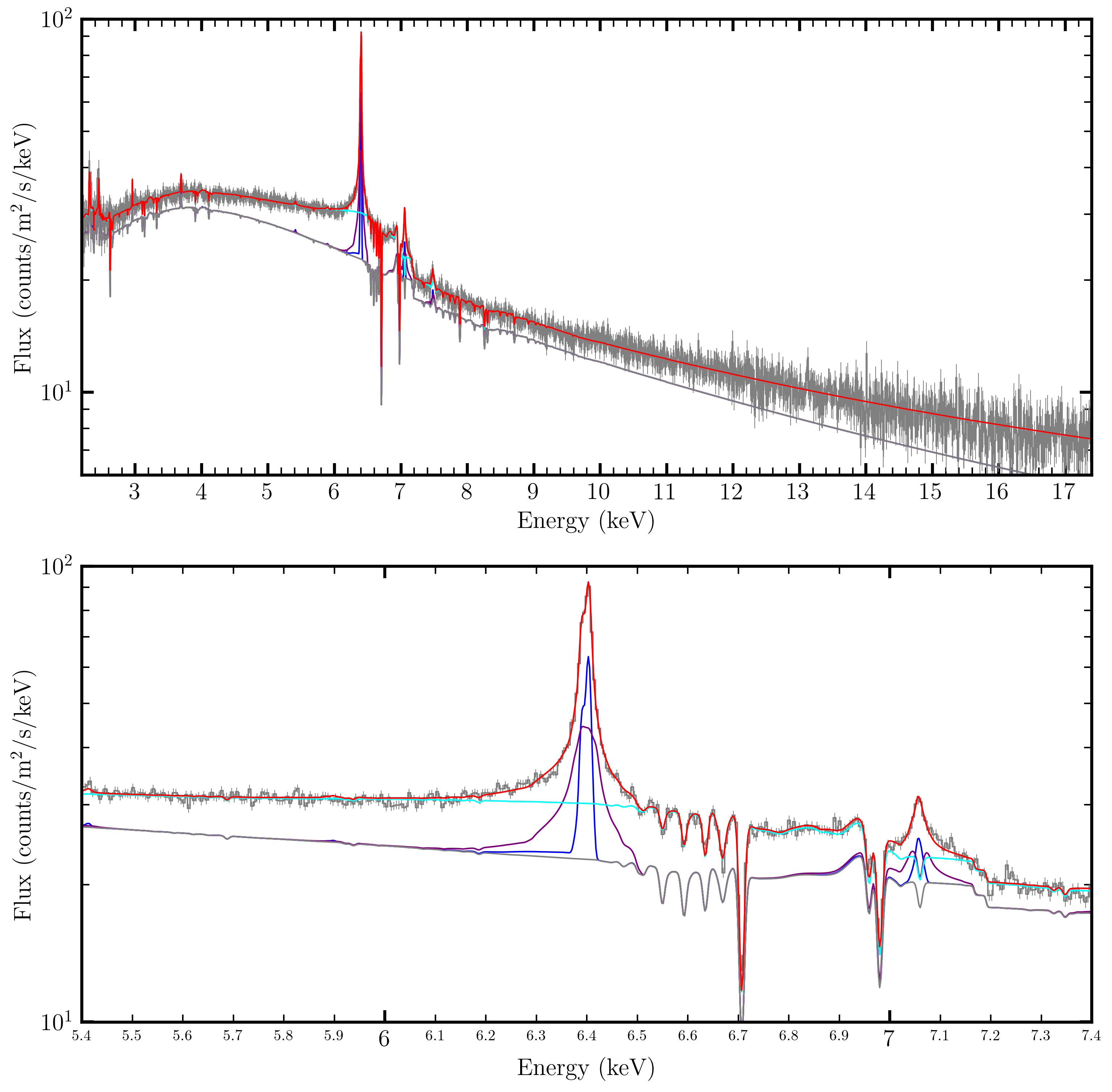}
    \caption{Top: The 0.9~Ms Resolve spectrum of NGC 4151 over the full fitted pass band.  Bottom: The spectrum in the Fe~K band. The data and models are plotted in the host frame.  In this figure, contributions to the Fe~K emission line structure from the inner disk (cyan), broad line region (purple), and torus (blue) are highlighted separately.  The direct power-law is shown in gray.  Note that reflection carries some continuum flux; deviations from the power-law are not only due to line flux.}
    \label{fig:broad_comps}
    \vspace{-0.5in}
\end{figure*}

\begin{figure*}
    \centering
    \includegraphics[width=1.0\textwidth]{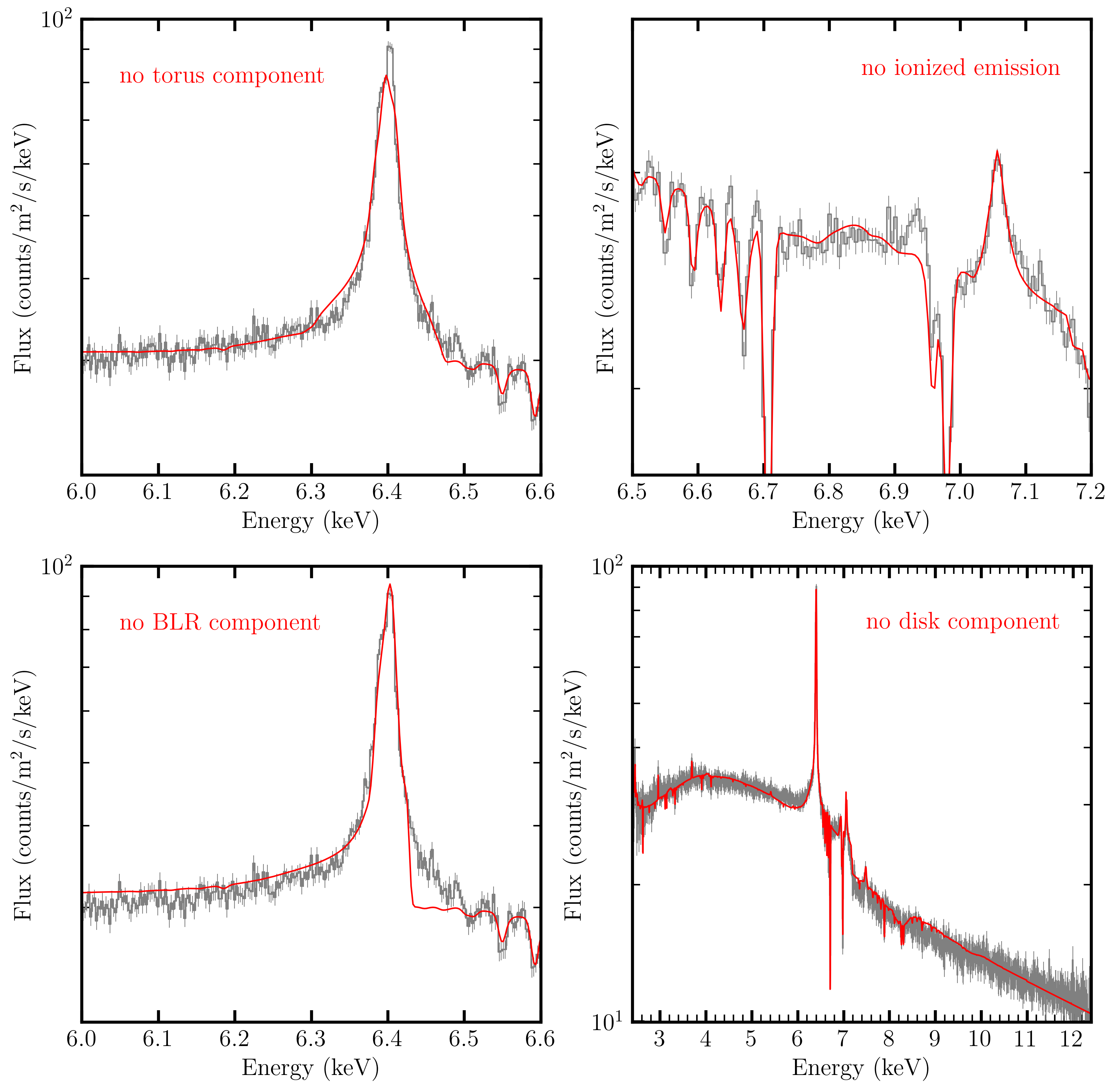}
    \caption{The 0.9~Ms Resolve spectrum of NGC 4151, shown after removing the emission component indicated in each panel, and re-fitting while allowing all of the other model parameters to vary.  Top left: the model in red lacks a contribution from the distant torus, and fails to fit the narrow core of the line. Bottom left: the model in red lacks a contribution from the BLR, and fails to match the wings of the Fe~K$_{\alpha}$ line.  Top right: the model in red lacks re-emission from the most ionized warm absorber, and fails to fit the P Cygni line profile. Bottom right: the model lacks relativistic reflection from the inner disk, and fails to fit the data in the Fe~K band and more broadly.}
    \label{fig:emis}
\end{figure*}

\begin{figure*}
    \centering
    \includegraphics[width=1.0\textwidth]{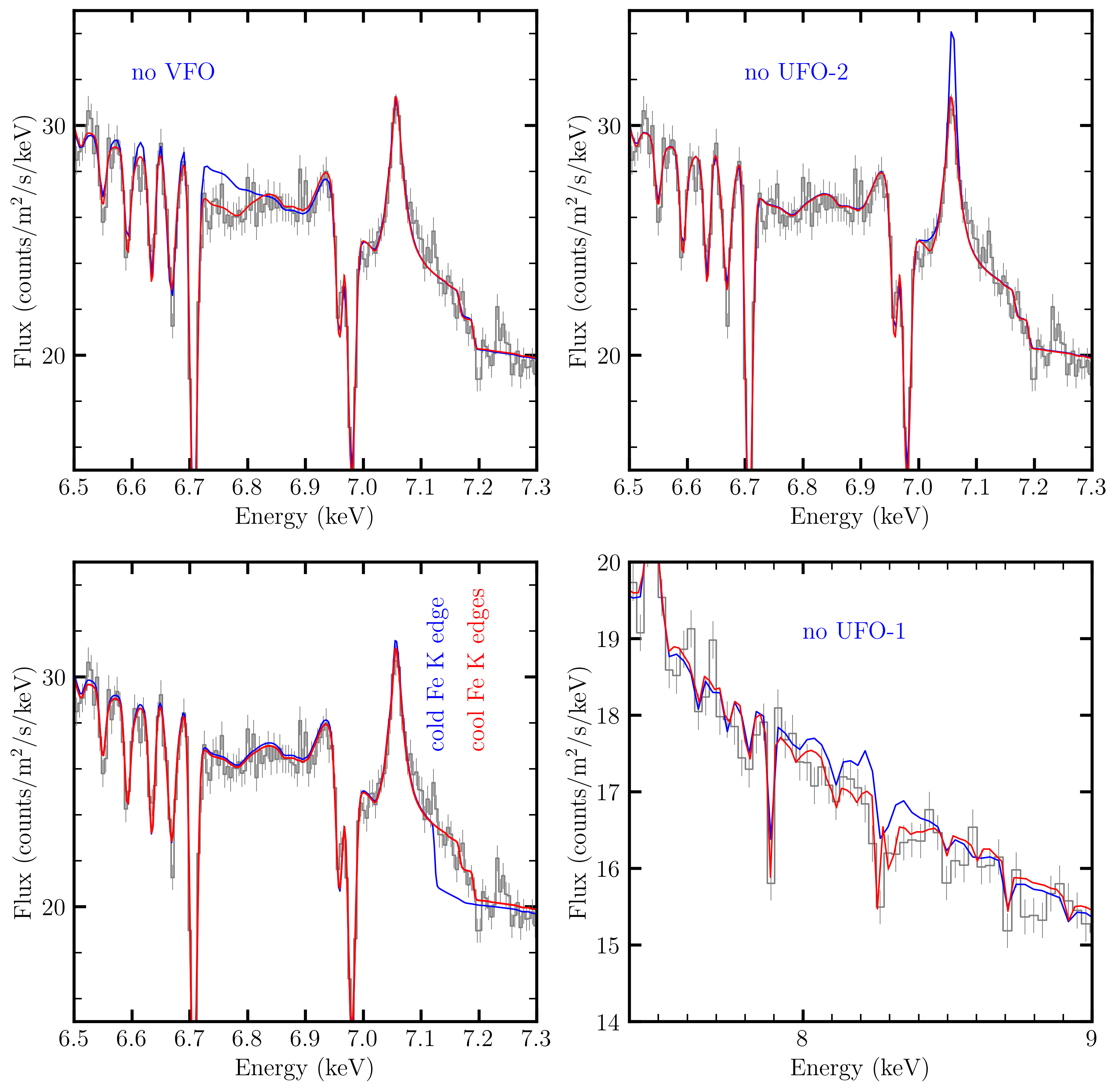}
    \caption{The 0.9~Ms Resolve spectrum of NGC 4151.  In each panel, the best-fit model is shown in red, whereas the model in blue depicts the model after a given absorption component is removed 
    or altered and the data are fit again.  Top left: the model in blue excludes the very fast outflow component, and fails to fit the Fe XXV absorption between 6.7--6.8 keV.  Top right: the model in blue lacks the second, slower UFO component, which partially absorbs the Fe~K$_{\beta}$ line.  Bottom right: the model in blue lacks the $-v = 0.17c$ UFO component, which places a broad Fe~XXVI absorption line at 8.2~keV.  Bottom left: the model in blue describes the Fe K edge in terms of cold, neutral gas, whereas the best-fit model requires cool $kT = 5.5^{+0.4}_{-0.3}$~eV gas.}
    \label{fig:emis}
\end{figure*}



\end{document}